\documentclass{style}
\usepackage{graphicx}
\usepackage[utf8]{inputenc}
\usepackage[T1]{fontenc}
\usepackage{lmodern}
\usepackage{graphicx}
\usepackage{dcolumn}
\usepackage{bm}
\usepackage{amsmath}
\usepackage{caption}[2008/07/24]
\newcommand{\beq}{\begin{equation}}
\newcommand{\eeq}{\end{equation}}
\newcommand{\beqa}{\begin{eqnarray}}
\newcommand{\eeqa}{\end{eqnarray}}
\renewcommand{\thefootnote}{\#\arabic{footnote}}

\newcommand{\mpi}{M_{\pi}}

\newcommand{\Order}{\mathcal{O}}

\newcommand{\mN}{m_N}
\newcommand{\sm}{s_\text{m}}
\newcommand{\MeV}{\,\text{MeV}}
\newcommand{\GeV}{\,\text{GeV}}

\newcommand{\fet}[1]{\mbox{\boldmath $#1$}}
\newcommand{\nn}{\nonumber \\ }

\begin{document}

\title{The long and winding road from chiral effective Lagrangians to nuclear structure}
\author{
  Ulf-G Mei{\ss}ner \email{meissner@hiskp.uni-bonn.de} \\
  {\it Universit\"at Bonn, Helmholtz-Institut f\"ur Strahlen- und Kernphysik and}\\ 
  {\it Bethe Center for Theoretical Physics,
   D-53115 Bonn, Germany}\\and\\
  {\it Forschungszentrum J\"ulich, Institut f\"ur Kernphysik, Institute for Advanced Simulation}\\
  {\it and J\"ulich Center for Hadron Physics, D-52415 J\"ulich, Germany}
}


\date{}

\maketitle

\begin{abstract}
  I review the chiral dynamics of nuclear physics. In the first part, I discuss the new developments in the construction of the forces
  between two, three and four nucleons which have been partly carried  out to fifth order in the chiral expansion. It is also shown
  that based on these forces in conjunction with the estimation of the corresponding theoretical uncertainties, the need for three-nucleon
  forces in few nucleon systems can be unambiguously established. I also introduce the lattice formulation of these forces, which
  allow for truly {\em ab initio} calculations of nuclear structure and reactions. I present some pertinent results of the nuclear lattice approach.
  Finally, I discuss how few-nucleon systems and nuclei can be used to explore symmetries and physics within and beyond the Standard Model.
\end{abstract}


\section{Introduction}
\label{sec:intro}

This contribution to celebrate the 40$^{\rm th}$ anniversary of the Nobel prize to the nuclear structure investigations of Bohr, Mottelson and Rainwater\footnote{To my
opinion my beloved teacher, Gerald (Gerry) E. Brown has also made contributions to nuclear physics  worthy of the Nobel prize. I therefore dedicate
this paper to his memory.} reviews some
work that firmly links nuclear physics to the gauge theory of the strong interactions, Quantum Chromodynamics (QCD),  by the exploration of the
symmetries of QCD and their realization. This is arguably the most important development in nuclear physics since many decades, as it puts 
nuclear physics on a very different level of rigor and precision than was possible before. The main ingredient in such an approach is the concept
of an effective Lagrangian, that was championed for the strong interactions by Noble laureate Steven Weinberg \cite{Weinberg:1978kz}
and by Gasser and Leutwyler~\cite{Gasser:1983yg,Gasser:1984gg}. 
In such an approach, one
is able to perform a systematic expansion in a small parameter, typically some soft external momentum or a small mass divided by a hard (large) scale.
Thus scale separation is an important ingredient, and that is exactly what the spectrum of QCD for the light flavors up, down and strange exhibits.
Further, such an effective Lagrangian approach allows to estimate the uncertainty  of any given calculation, an absolute must for any serious
theoretical approach. Or stated more bluntly: A theoretical calculation that does not give an uncertainty is as good as any random number. 
In nuclear physics, matters are, however, a bit more complicated, as the very small binding energies or binding momenta seem to restrict the
applicability of the effective Lagrangian approach to a very small range of energies or momenta. Again, it was Weinberg~\cite{Weinberg:1990rz,Weinberg:1991um}, who laid out the
framework to overcome these obstacles. 
His approach has been criticized by many, but so far only within the so-called Weinberg power counting
scheme to be explained in detail below, nuclear structure questions can be addressed rigorously. Under certain circumstances, the scale separation
can be used to set up other effective theories, such as the pionless nuclear effective  field theory (for reviews, see e.g. 
Refs.~\cite{Bedaque:2002mn,Braaten:2004rn})  or the effective theory for halo nuclei \cite{Bertulani:2002sz}. These, however,
will not be discussed here. Another more direct path from QCD is the application of lattice QCD to nuclear systems. Such calculations
have, however, to overcome severe obstacles, and will not be a precision tool in the next few years. Still, lots of progress is made in that approach,
as witnessed e.g. by recent calculations of light nuclei and hyper-nuclei at large pion masses~\cite{Beane:2012vq},
of the  magnetic moments of nuclei~\cite{Beane:2014ora} or trying to construct a nuclear 
potential~\cite{Aoki:2012tk}.

This paper is organized as follows. In Sec.~\ref{sec:qcd} I review the essentials of QCD, with particular emphasis on the sector of the light quarks.
This is further elaborated in Sec.~\ref{sec:chisym}, where chiral symmetry and its various variants of breaking are discussed. There is  also a short
discussion of the broken U(1)$_A$ symmetry of QCD, which can be explored to test physics beyond the Standard Model. The next section~\ref{sec:Leff}
introduces the concept of the effective Lagrangian and the machinery related to it, in particular the power counting, the so-called low-energy
constants and issues related to renormalization. An important ingredient to construct nuclear forces is pion-nucleon scattering, which is discussed
in Sec.~\ref{sec:piN}, including very recent results from Roy-Steiner equations matched to chiral perturbation theory, the effective field theory
(EFT) of QCD. Armed with that, nuclear forces are discussed in Sec.~\ref{sec:forces} featuring the most recent results at fifth order in the chiral
expansion for two-nucleon forces, the estimation of theoretical uncertainties and the status of three- and four-nucleon forces.  To tackle nuclei,
one can either use these forces in connection with more conventional many-body approaches (shell model, coupled-cluster methods and so on)
or discretize space-time and use this lattice to perform Monte Carlo simulations of nuclei. It is this latter approach which will be described in Sec.~\ref{sec:latt}, and
assorted results will be presented in Sec.~\ref{sec:res}.  One of the nice features of this novel framework is that it allows to investigate the behavior of nuclear
structure and reactions under variations of the fundamental parameters, e.g. the light quark masses and the electromagnetic fine-structure constant.
This is discussed in Sec.~\ref{sec:anthro} together with its consequences for our anthropic view of the Universe. The role of nuclei as precision
laboratories to explore symmetries within and beyond the Standard Model is reviewed in Sec.~\ref{sec:precision}.  The final section~\ref{sec:out}
gives some perspectives and outlines future research in this exciting area of physics.


\section{QCD}
\label{sec:qcd}

QCD is a fascinating theory based on a local, non-abelian  SU(3)$_{\rm color}$ symmetry. It embodies all of strong interaction 
physics essentially in one simple line\footnote{This form  is a bit  simplified, as discussed later on.}
\beq
\label{eq:QCD}
{\cal L}_{\rm QCD} = -\frac{1}{2g^2} {\rm Tr} \left( G_{\mu\nu}G^{\mu\nu}
\right) + \bar \psi \,\left( i \gamma^\mu D_\mu - \ {\mathcal M} \right) \psi~,
\eeq
where $G_{\mu\nu}$ is the gluon field strength tensor, and $\psi$ is a spinor that includes the six quark flavors up, down, strange,
charm, bottom and top. The forces are mediated by gluons that couple to the
color not made explicit in Eq.~(\ref{eq:QCD})  and  we have absorbed the gauge coupling in the definition of the gluon
field. As the gluons carry color, they can interact with themselves through 3- and 4-point vertices.  The gauge-covariant
derivate $D_\mu$ generates the quark-gluon coupling. It is important to realize that QCD can be split into two sectors, one referring to the 
light quarks $u$, $d$, and $s$ and the other one is given by the heavy quarks $c$ and $b$\footnote{The top quark decays too quickly to 
form any strongly interacting particle.}.
Here, light means that the current quark mass is much smaller than the typical scale of QCD, $\Lambda_{\rm QCD} \simeq 250\,$MeV,
that can be inferred from the running of the strong coupling constant $\alpha_s = g^2/(4\pi)$, whereas the mass of the heavy
quarks is much larger than this scale.
In the light quark sector, one can rewrite the QCD Lagrangian in terms of the three-component vector $q$ that collects the light
quark fields, $q^T (x) = \left( u(x), d(x), s(x)\right)$.  Here, the mass term can be considered as a perturbation, and to first order,
one can completely ignore the mass term. This is what Heiri Leutwyler calls a ``theoretical paradise''  \cite{Leutwyler:2012ax}
as one is dealing with a 
theory that has not a single tunable parameter as $\Lambda_{\rm QCD}$ is generated by dimensional transmutation. As it is well known,
massless fermions exhibit a chiral symmetry, that is, the theory exists in two copies, one for the left- and the other for the right-handed
fields.  There is no  interaction between these two almost identical worlds.  More on that in the next chapter. The heavy quarks
are commonly collected in the doublet $Q^T(x) = \left( c(x), b(x)\right)$, so that the leading order Lagrangian of the heavy quark
effective theory, in which the large masses $m_c$ and $m_b$ has been transformed into a string of $1/m_Q$ suppressed terms,
simply takes the form ${\cal L}_0 = \bar Q \, (i v\cdot D)\, Q$, with $v_\mu$ the four-velocity of the heavy quark. This form obviously exhibits
a SU(2)$_{\rm spin}$ symmetry as well as an SU(2)$_{\rm flavor}$ symmetry, combined in the SU(4) spin-flavor symmetry, as 
 ${\cal L}_0$ neither depends on the spin of the heavy quark nor on its mass. These symmetries are, of course, broken at
 next-to-leading order  from corrections to the kinetic energy term and the chromo-magnetic Pauli interaction. Nothing more will be said
 here on this intriguing part of the theory. The quarks and gluons are confined within hadrons, the strongly interacting particles.
 Most hadrons are simple mesons (quark-antiquark states) or baryons (three quark states), but as of today some tetraquark
 and may be even pentaquark states have been established. It is still not understood why QCD mostly generates states of the
 simplest types and also the expected glueballs, that are made of nothing but glue, have been elusive. However, it is basically known
 how the massive hadrons acquire their mass. Bound states made of heavy quarks are essentially slow moving objects, with their
 mass largely given by the masses of the quarks they are made of. For the hadrons  made of light quarks (with the exception of
 the Goldstone bosons to be discussed below), most of the mass is generated by the gluon field energy, beautifully realizing Wheeler's
 notion of ``mass without mass'' \cite{Wheeler}.
 More formally, this can be understood from the breaking of the dilatation current of massless classical QCD,
 the so-called trace anomaly~\cite{Collins:1976yq}. As one example, the nucleon mass can be written as follows:
 \beqa
 & m_N \bar u (p) u(p) = \langle N(p) | \theta_\mu^\mu| N(p)\rangle  =& \nonumber\\
 &  \langle N(p) | \frac{\beta_{\rm QCD}}{2 g}
 G_{\mu\nu}^a G^{\mu\nu}_a + m_u \bar u u + m_d \bar d d + m_s \bar s s |N(p)\rangle~, & \nonumber \\ 
 &&
 \eeqa
 where $\beta_{\rm QCD}$ is the QCD $\beta$-function and $\theta_\mu^\mu$ the trace of the energy-momentum tensor. 
 The first term gives the field energy contribution, the
 terms proportional to $m_u$ and $m_d$ can be related to the pion-nucleon sigma term and the last one to the
 strangeness content. As discussed below, these sum up to about 110~MeV (with sizeable uncertainty for the strangeness
 contribution), so that
 the bulk of the mass is indeed coming from the gluon field energy. This scenario is also consistent with recent 
 lattice QCD calculations of hadron masses at physical quark masses.
 
 As noted in the footnote~\#2, there is a bit more to QCD than just given in Eq.~(\ref{eq:QCD}). I refer to the so-called
$\theta$-term of QCD, that is a consequence of the non-trivial vacuum structure of QCD and the anomalous breaking
of the U(1)$_A$ symmetry, given by
\begin{equation}
{\cal L}_\theta = -\displaystyle\frac{\theta}{64 \pi^2} \,
\epsilon^{\mu\nu\rho\sigma} G^a_{\mu\nu}G^a_{\rho\sigma}\,,
\end{equation}
where $a =1, \ldots,8$ are color indices. The $\theta$-term can be rotated into the quark mass matrix, and as long as
one of the current quark masses vanishes, so does the effect of $\theta$. However, nature
does not seem to pull this option as all quark masses are non-vanishing. The $\tilde GG$ term, with
$\tilde G^{\alpha\beta}  \sim \epsilon^{\alpha\beta\gamma\delta}G_{\gamma\delta}$, clearly
has odd properties as it is proportional to the product of the chromo-electric and the chromo-magnetic
fields and thus is odd under CP (charge conjugation times parity transformation), and assuming CPT to be an exact symmetry, 
also under T (time reversal, or, more correctly, motion reversal, as Cecilia Jarlskog  often points  out, see e.g.~\cite{Jarlskog} ).  Consequently,
hadrons can acquire permanent electric dipole moments (EDMs)\footnote{Such a permanent  EDM should not be confused 
with an {\em induced} dipole moment. Such induced dipole moments are typically of the size or smaller than the volume of the hadron under
consideration.} that can interact with external electric fields. 
Measurements of the upper limit of electric dipole moment of the neutron poses
a stringent limit on the value of $\theta$~\cite{Crewther:1979pi},
where the latest determination gives $|\theta| < 7.6 \cdot 10^{-11}$~\cite{Guo:2015tla}.


\section{Chiral symmetry}
\label{sec:chisym}

First, let me introduce the concept of chiral 
symmetry. Consider a theory of massless fermions,  
\beq
{\mathcal L} = i \bar \psi \gamma_\mu \partial^\mu \psi~.
\eeq
Such a theory possesses a {\em chiral symmetry}. To see
this, perform a left/right~(L/R)-decomposition of the spin-1/2 field
\beq\label{LRsplit}
\psi = \frac{1}{2}(1-\gamma_5) \psi + 
 \frac{1}{2}(1+\gamma_5) \psi = P_L \psi + P_R \psi = \psi_L + \psi_R~,
\eeq
using the  projection operators $P_{L/R}$, that obey
$P_L^2 = P_L,~ P_R^2 = P_R,~ P_L \cdot P_R = 0,~ P_L +
P_R = {\rm 1\hspace{-0.65ex}I}$. The $\psi_{L/R}$ are
helicity eigenstates
\beq
\frac{1}{2} \hat{h} \psi_{L/R} = \pm 
\frac{1}{2}  \psi_{L/R}~,~~~ \hat h 
= \displaystyle\frac{\vec\sigma \cdot \vec p}{|\vec{p\,}|}~,
\eeq
where $\vec{p}$ denotes the fermion momentum and $\vec\sigma$ are
the Pauli spin matrices. In terms of the left- and right-handed
fields, the Lagrangian takes the from
\beq
{\mathcal L} = 
i \bar \psi_L \gamma_\mu \partial^\mu \psi_L +
i \bar \psi_R \gamma_\mu \partial^\mu \psi_R~,
\eeq
which means that the L/R fields do {\em not} interact and,
by use of Noether's theorem, one has  conserved L/R currents.
We note that a fermion mass term breaks chiral symmetry, as 
such a  term mixes the left- and right-handed components, 
$\bar \psi {\cal M} \psi = \bar\psi_R {\cal M} \psi_L 
+ \bar \psi_L {\cal M} \psi_R$. Physically, this is easy to
understand. While massless fermions move with the speed of light,
this is no longer the case for massive fermions. Thus, for a massive
fermion with a given handedness in a certain frame, one can always
find a boost such that the sign of $\vec\sigma \cdot \vec p$ changes.
If the mass term is sufficiently small (where ``small'' depends on 
other scales in the theory), one can treat this {\em explicit}
chiral symmetry breaking in perturbation theory and speaks of
an {\em approximate} chiral symmetry -- more on that later. 

In many fields of physics, broken symmetries play a special role.
An intriguing phenomenon is {\em spontaneous} symmetry breaking,
which means that the ground state of a theory shares a lesser symmetry
than the corresponding Lagrangian or Hamiltonian. A key ingredient 
in this context is Goldstone's theorem 
\cite{Goldstone:1961eq,Goldstone:1962es}: 
To every generator of a spontaneously broken symmetry corresponds a 
massless excitation of the vacuum. 
This can be understood in a nut-shell (ignoring subtleties like
the normalization of states and alike - the argument also goes through in 
a more rigorous formulation). Let ${\cal H}$ be some Hamiltonian that is invariant
under some charges $Q^i$, i.e. $[{\cal H}, Q^i] = 0$, with $i = 1, \ldots, n$.
Assume further that $m$ of these charges ($m \leq n$) do not annihilate the
vacuum, that is $Q^j |0 \rangle \neq 0$ for $j =1, \ldots, m$.  Define a
single-particle state via $|\psi\rangle = Q^j |0 \rangle$.   This is an
energy eigenstate with eigenvalue zero, since $H |\psi\rangle  = H
Q^j|0\rangle=  Q^j H
|0\rangle = 0$. Thus, $|\psi\rangle$ is a single-particle state with $E =
\vec{p} = 0$, i.e. a massless excitation of the vacuum. These states are the
{\em Goldstone bosons}, collectively denoted as pions $\pi(x)$ in what follows.
Through the corresponding symmetry current the Goldstone bosons couple 
directly to the vacuum,
\beq\label{GBME}
\langle 0 | J^0 (0) | \pi \rangle \neq 0~.
\eeq
In fact, the non-vanishing of this matrix element is a {\it necessary and
sufficient} condition for spontaneous symmetry breaking. 

Another important property of Goldstone bosons is the derivative nature of their
coupling to themselves or matter fields. Again, in a hand-waving fashion, this
can be understood easily. As above, one can repeat the operation of acting
with the non-conserved charge  $Q^j$ on the vacuum state $k$ times, thus 
generating a state of $k$ Goldstone bosons that is degenerate with the
vacuum. Assume now that the interactions between the Goldstone bosons is
not vanishing at zero momentum. Then, the ground state ceases to be degenerate
with the $k$ Goldstone boson state, thus the assumption must be incorrect.
Of course, this argument can also be made rigorous.
In the following, the derivative nature of the pion couplings will play an 
important role.

Let us now consider three-flavor QCD with up, down, and strange quarks.
As far as the strong interactions
are concerned, the different quarks $u,d,s$ have identical properties, except
for their masses. The quark masses are free parameters in QCD - the theory
can be formulated for any value of the quark masses. In fact, light quark QCD
can be well approximated by a fictitious world of massless quarks.  Remarkably,
this theory contains no adjustable parameter - the gauge coupling $g$ merely
sets the scale for the renormalization group invariant scale $\Lambda_{\rm
  QCD}$. The Lagrangian of massless QCD is invariant under separate unitary
global transformations of the L/R quark fields, 
\beq
q_I \to V_I q_I~, \quad V_I \in U(3)~, \quad I = L,R~,
\eeq
leading to $3^2 =9$ conserved left- and $9$ conserved right-handed 
currents by virtue of Noether's theorem. These can be expressed in terms of 
vector ($V \sim L + R$) and  axial-vector ($A \sim L - R$) currents
\beq\label{conscurr}
V_0^\mu \,(A^\mu_0) = \bar q \, \gamma^\mu \, (\gamma_5)\, q~, ~~
V_\mu^a \,(A_\mu^a)= \bar q \, \gamma^\mu (\gamma_5)\frac{\lambda_a}{2} \, q~,
\eeq
Here, $a = 1, \ldots ,8$, and the $\lambda_a$ are Gell-Mann's SU(3)
flavor matrices. We remark that the singlet axial current is anomalous,
and thus not conserved. The actual symmetry group of massless QCD
is generated by the charges of the conserved currents, it is
$G_0 = {\rm SU(3)}_R \times {\rm SU(3)}_L \times {\rm U}(1)_V$. The U(1)$_V$ subgroup of 
$G_0$ generates conserved baryon number since the isosinglet vector 
current counts the number of quarks minus antiquarks in a hadron. 
The remaining group SU(3)$_R \times$\,SU(3)$_L$ is often
referred to as chiral SU(3). In what follows, we will mostly  consider the light
$u$ and $d$ quarks only (with the strange quark mass fixed at its
physical value). In that case, one speaks of chiral SU(2) and
must replace the generators in Eq.~(\ref{conscurr}) by the Pauli-matrices.

The chiral symmetry is a symmetry of the Lagrangian of QCD but not of the
ground state or the particle spectrum -- to describe the strong interactions
in nature, it is crucial that chiral symmetry is spontaneously broken. This
can be most easily seen from the fact that hadrons do not appear in parity
doublets. If chiral symmetry were exact, from any hadron one could generate
by virtue of an axial transformation another state of exactly the same 
quantum numbers except of opposite parity. The spontaneous symmetry breaking
leads to  the formation of a quark condensate in the vacuum  
$\langle 0 | \bar q q|0\rangle =\langle 0  | \bar q_L q_R + \bar q_R q_L|0\rangle$, 
thus connecting the left- with the right-handed
quarks. In the absence of quark masses this expectation value
is flavor-independent: $\langle 0 | \bar u u|0\rangle = 
\langle 0 | \bar d d|0\rangle = \langle 0 | \bar q q|0\rangle$. 
More precisely, the vacuum is only invariant under the subgroup of 
vector rotations times the baryon number current, $H_0 = {\rm SU(3)}_V \times
{\rm U(1)}_V$. This is the generally accepted picture that is supported by general
arguments \cite{Vafa:1983tf} as well as lattice simulations of QCD.
In fact, the vacuum expectation value of the quark condensate is only one
of the many possible order parameters characterizing the spontaneous symmetry
violation - all operators that share the invariance properties of the
vacuum (Lorentz invariance, parity, invariance under SU(3)$_V$
transformations) qualify as order parameters. The quark condensate
nevertheless enjoys a special role, it can be shown to be related to the
density of small eigenvalues of the QCD Dirac operator 
(see \cite{Banks:1979yr} and further discussions  in \cite{Leutwyler:1992yt,Stern:1998dy}),
\beq
\lim_{{\cal M} \to 0} \langle 0 | \bar q q|0\rangle = - \pi \, \rho(0)~.
\eeq
For free fields, $\rho (\lambda) \sim \lambda^3$ near $\lambda = 0$. Only if
the eigenvalues accumulate near zero does one obtains a non-vanishing condensate.
This scenario is indeed supported by lattice simulations and many model studies
involving topological objects like instantons or monopoles.

In QCD, we have
eight (three) Goldstone bosons for SU(3) (SU(2)) with spin zero and 
negative parity -- the latter property is a consequence that these Goldstone 
bosons are generated by applying the axial charges on the vacuum. The
dimensionfull scale associated with the matrix element Eq.~(\ref{GBME}) 
is the pion decay constant (in the chiral limit)
\beq\label{GBM}
\langle 0|A^a_\mu(0)|\pi^b(p)\rangle = i \delta^{ab} F p_\mu~,
\eeq
which is a fundamental mass scale of low-energy QCD.
In the world of massless quarks, the value of $F$ differs from the
physical value by terms proportional to the quark masses, to be
introduced later, $F_\pi = F [1 + {\cal O}({\cal M})]$. The physical
value of $F_\pi$ is $92.2\,$MeV, determined from pion decay, $\pi\to \nu\mu$.

Of course, in QCD the quark masses are not exactly zero. The quark mass term leads
to the so-called {\em explicit chiral symmetry breaking}. Consequently, the
vector and axial-vector currents are no longer conserved (with the exception
of the baryon number current)
\beq\label{div}
\partial_\mu V_a^\mu = \dfrac{1}{2} i \bar q \, [{\cal M},\lambda_a ]\, q~, \quad
\partial_\mu A_a^\mu = \dfrac{1}{2} i \bar q \, \{{\cal M},\lambda_a \} \, \gamma_5\, q~.
\eeq
However, the consequences of the spontaneous symmetry violation  can still be
analyzed systematically because the quark masses are {\em small}. QCD
possesses what is called an approximate chiral symmetry. In that case, the mass spectrum
of the unperturbed Hamiltonian and the one including the quark masses can not be
significantly different. Stated differently, the effects of the explicit symmetry
breaking can be analyzed in perturbation theory. This perturbation generates
the remarkable mass gap of the theory - the pions (and, to a lesser extent,
the kaons and the eta) are much lighter than all other hadrons.  To be more
specific, consider chiral SU(2). The second formula of Eq.~(\ref{div}) is
nothing but a Ward-identity  that relates the axial current $A^\mu =  \bar d
\gamma^\mu \gamma_5 u$ with the pseudoscalar density $P = \bar d i \gamma_5
u$,
\beq
\partial_\mu A^\mu = (m_u + m_d)\, P~.
\eeq
Taking on-shell pion matrix elements of this Ward-identity, one arrives at
\beq\label{pimass}
M_\pi^2 = (m_u + m_d) \frac{G_\pi}{F_\pi}~,
\eeq
where the coupling $G_\pi$ is given by $\langle 0 | P(0)| \pi(p)\rangle =
G_\pi $. This equation leads to some intriguing consequences: In
the chiral limit, the pion mass is exactly zero - in accordance with Goldstone's
theorem. More precisely, the ratio $G_\pi /F_\pi$ is a constant in the chiral
limit and the pion mass grows as $\sqrt{m_u+m_d}$ as the quark masses are
turned on.

There is even further symmetry related to the quark mass term. It is observed
that hadrons appear in isospin multiplets, characterized by very tiny
splittings of the order of a few MeV. These are generated by the small
quark mass difference $m_u -m_d$ (small with respect to the
typical hadronic mass scale of a few hundred MeV)
and also by electromagnetic effects of the
same size (with the notable exception of the charged to neutral pion mass
difference that is almost entirely of electromagnetic origin). This can be
made more precise: For $m_u = m_d$, QCD is  invariant under SU(2) isospin 
transformations: 
\beqa 
q \to q'  &=& U q~,~~ q = \left(\begin{array}{cc} u \\ d\end{array}\right)~,\nonumber \\
U &=& \left(\begin{array}{cc} a^* &  b^* \\ -b & a\end{array}\right)~,
~~~|a|^2+|b|^2 = 1~.
\eeqa
In this limit, up and down quarks can not be disentangled as far as the
strong interactions are concerned.  Rewriting of the QCD quark mass term
allows to make the strong isospin violation explicit:
\beqa
{\cal H}_{\rm QCD}^{\rm SB} &=& m_u \,\bar u u + m_d
  \,\bar  d d \nonumber \\ &=& \dfrac{m_u+m_d}{2}(\bar u u + \bar d d)
              + \dfrac{m_u-m_d}{2}(\bar u u - \bar d d)~,\nonumber\\
\eeqa
where the first (second) term is an isoscalar (isovector). Extending 
these considerations to SU(3), one arrives at the eightfold way of
Gell-Mann and Ne'eman \cite{GMNbook}
that played a decisive role in our understanding
of the quark structure of the hadrons. The SU(3) flavor symmetry
is also an approximate one, but the breaking is much stronger than it is
the case for isospin. From this, one can directly infer that the quark mass
difference $m_s - m_d$ must be much bigger than $m_d -m_u$.

There is one further source of symmetry breaking, which is best understood
in terms of the path integral representation of QCD. The effective
action contains an integral over the quark fields that can be expressed
in terms of the so-called fermion determinant. Invariance of the theory
under chiral transformations not only requires the action to be left
invariant, but also the fermion measure \cite{Fujikawa:1983bg}. Symbolically,
\beq
\int [d\bar q][dq] \ldots \to |{\mathcal J}|\int [d\bar q'][dq'] \ldots
\eeq   
If the Jacobian is not equal to one, $|{\mathcal J}| \neq 1$,
one encounters an {\it anomaly}. We already encountered one
example, the $\theta$-term of the QCD Lagrangian.
Of course, such a statement has to be 
made more precise since the path  integral requires regularization and 
renormalization, still it captures the essence of the chiral anomalies
of QCD. One can show in general that certain 3-, 4-, and 5-point
functions with an odd number of external axial-vector sources are 
anomalous. As particular examples we mention the famous triangle
anomalies of Adler, Bell and Jackiw and the divergence of the
singlet axial current,
\beq
\partial_\mu (\bar q \gamma^\mu \gamma_5 q) = 2iq m \gamma_5 q +
\frac{N_f}{8\pi} G_{\mu\nu}^a \tilde{G}^{\mu\nu, a}~,
\eeq
that is related to the generation of the $\eta '$  mass. There are
many interesting aspects of anomalies in the context of QCD and chiral
perturbation theory~\cite{Bijnens:1993xi}. In what follows, we will
consider some consequences of the $\theta$-term in QCD.


\section{Effective Lagrangian}
\label{sec:Leff}

 To deal with systems that exhibit scale separation, one considers a properly formulated
 {\em effective Lagrangian} that shares the same symmetries as the underlying theory
 (like in our case QCD) but is formulated in terms of the pertinent asymptotic hadronic fields,
 here pions and nucleons. To keep matters simple, let us first consider pions only.
As the pions are  Goldstone bosons, their interactions are of derivative nature. This 
allows to formulate an EFT at low energies/momenta, as derivatives can be translated
into small momenta. Such an EFT is necessarily non-renormalizable,
as one can write down an infinite tower of terms with increasing number
of derivatives consistent with the underlying symmetries, in particular
chiral symmetry. Consequently, such an EFT can only be applied for
momenta and masses (setting the ``soft'' scale) that are small compared
to masses of the particles not considered (setting the '``hard'' scale).
For the case at hand, the hard scale is of the order of 1~GeV\footnote{This is
only a rough estimate as in some channels resonances (the heavy degrees of freedom)
might show up earlier, as it is the case for S-wave, isospin zero pion-pion interactions,
that feel the low-lying but broad $f_0(500)$ resonance.}. 
Let me  now show that there is a hierarchy of terms that allows one to make
precise predictions with a quantifiable theoretical order. This scheme
is called {\em power counting}. To be precise, consider
an effective Lagrangian
\beq
{\cal L}_{\rm eff} = \sum_d {\cal L}^{(d)}~,
\eeq
where $d$ is supposed to be bounded from below. 
For
interacting Goldstone bosons, $d \ge 2$, and $d$ is even for pionic
interactions due to Lorentz invariance or parity. The pion 
propagator is $D(q) = {i}/({q^2-M^2_\pi})$, with $M_\pi$ the
pion mass. Consider now  an $L$-loop diagram with $I$ internal lines 
and $V_d$ vertices of order $d$. The corresponding amplitude
scales as following
\beq\label{amp}
Amp \propto \displaystyle\int (d^4q)^L \, \displaystyle\frac{1}{(q^2)^I} \prod\limits_d
(q^d)^{V_d}~,
\eeq
where we only count powers of momenta. Now let $Amp \sim q^\nu$, therefore
using Eq.~(\ref{amp}) gives $\nu = 4L -2I + \sum_d d V_d$.
Topology relates the number of loops to the number of internal
lines and vertices as  $L = I - \sum_d V_d + 1$, so that we can
eliminate $I$ and arrive at the compact formula~\cite{Weinberg:1978kz}
\beq\label{pc}
\nu = 2 + 2L + \sum\limits_d V_d (d-2)~.
\eeq
The consequences of this simple formula are far-reaching. To lowest order (LO),
one has to consider only graphs with $d=2$ and $L=0$, which are tree
diagrams. Explicit symmetry breaking is also included as the quark mass counts
as two powers of $q$, cf. Eq.~(\ref{pimass}).  This LO contribution is nothing 
but the current algebra result, which can also be obtained with different - 
though less elegant - methods. However, Eq.~(\ref{pc})  tells us how to
systematically construct corrections to this. At next-to-leading order (NLO), 
one has one loop graphs $L=1$ built from the lowest oder interactions and also 
contact terms with $d=4$, that is higher derivative terms that are accompanied
by parameters, the so-called low-energy constants (LECs), that are not
constrained by the symmetries. These LECs must be fitted to data or can
eventually be obtained from lattice simulations, that allow to vary the 
quark masses and thus give much easier access to the operators that involve
powers of quark mass insertions or mixed terms involving quark masses and
derivatives. Space forbids to discuss this interesting field, I just refer
to the recent compilation in Ref.~\cite{Aoki:2013ldr}.
At next-to-next-to-leading order (NNLO), one has to consider two-loop
graphs with $d=2$ insertions, one-loop graphs with one $d=4$ insertion and
$d=6$ contact terms. Matter fields can also be included in this scheme. For
stable particles like the nucleon, this is pretty straightforward, the main
difference to the pion case is the appearance of operators with an odd number
of derivatives. For unstable states, the situation is more complicated, as one
has to account for the scales related to the decays. For example, in case of
the $\Delta(1232)$-resonance, one can set up a consistent power counting if
one considers the nucleon-delta mass difference as a small parameter. Here, I
will not further elaborate on these issues but rather refer to some early
related works on the $\Delta$ and vector mesons 
\cite{Hemmert:1997ye,Bruns:2004tj,Leupold:2008bp,Djukanovic:2009zn}.

\begin{figure}[t]
\begin{center}
\includegraphics[width=2cm,angle=270]{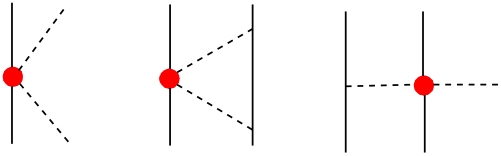}
\end{center}
\caption{\label{fig:ci}
The LECs $c_i$ (circles) in pion-nucleon scattering (left), the 
two-nucleon (NN) interaction (center) and the three-nucleon
(NNN) interaction (right).
}
\end{figure}

Coming back to chiral perturbation theory in this pure setting (considering pions
and possibly nucleons), it is worth to emphasize that the chiral Ward identities of 
QCD, that are faithfully obeyed in chiral perturbation theory
\cite{Leutwyler:1993iq,D'Hoker:1994ti},  connect a tower of different
processes involving various numbers of pions and external sources so
that fixing the low-energy constants  through a number of processes
allows one to make quite a number of testable predictions. Furthermore, as the
order increases, the number of LECs also increases, but again for a
specific process this is not prolific. The prime example is elastic
pion-pion scattering, which features four LECs at one-loop order but
only two new LECs appear at two loops - all other local two-loop
contributions to this reaction  merely
correspond to quark mass renormalizations of operators existing
already at one loop.  This is a more general phenomenon as  one can 
group the various operator
structures in two classes: The so-called {\em dynamical} operators
refer to terms with  derivatives on the hadronic fields (e.g. powers of
momenta)  and are independent of the quark
masses, whereas the so-called {\em symmetry-breakers} come with 
certain powers of quark mass insertions and thus vanish in the chiral
limit.  As stated before, lattice simulations that allow one to vary the quark
masses can be used efficiently to learn about this type of
operators. Let me come back to the interconnections between various processes 
in terms of the LECs. A particularly nice and appropriate  example that will 
be discussed in more detail later on is
related to the dimension-two couplings $c_i$ in the chiral effective
pion-nucleon Lagrangian, see Fig.~\ref{fig:ci} (for precise definitions
and further details, see the review \cite{Bernard:1995dp}).
The corresponding operators can e.g. be fixed in a fit to pion-nucleon
scattering data, see the left graph in Fig.~\ref{fig:ci}. This issue will be
taken up in some detail in the next section.
The same operators play not only an important role in
the two-pion exchange contribution to nucleon-nucleon scattering (middle graph in  Fig.~\ref{fig:ci}) but
also they give the longest range part of the three-nucleon forces (right graph in  Fig.~\ref{fig:ci}),
that are an important ingredient in the description of atomic nuclei
and their properties. In fact, there have also been attempts to
determine these couplings directly from nucleon-nucleon scattering
data, leading to values consistent with the ones determined from
pion-nucleon scattering. Furthermore, this clearly establishes the role
of pion-loop effects (see the middle graph in Fig.~\ref{fig:ci})
in nucleon-nucleon scattering beyond the
long-established tree-level pion exchange, already proposed by Yukawa
in 1935.  

One important issue to be discussed is unitarity. From the power
counting outlined above, it is obvious that imaginary parts of
scattering amplitudes or form factors are only generated at subleading
orders, or, more precisely, the one-loop graphs generate the leading
contributions to these. In general, this does not cause any problem,
with the exception of the strong pion-pion final state interactions related
to the low-lying and broad scalar $f_0(500)$ meson,
see e.g. Ref.~\cite{Meissner:1990kz} for an early discussion and the 
precise extraction of its properties from Roy equations~\cite{Caprini:2005zr}.
All this and more is nicely reviewed by Pelaez~\cite{Pelaez:2015qba}.
In fact, one can turn the argument around and use
analyticity and unitarity to calculate the  leading loop corrections
without ever working out a loop diagram -- the most famous examples
are  Lehmann's analysis of pion-pion scattering in 1972 \cite{Lehmann:1972kv}
and  Weinberg's general analysis of the structure of effective
Lagrangians \cite{Weinberg:1978kz}. A pedagogic introduction to the
relation between unitarity and CHPT can be found in
Ref.~\cite{Bernard:2006gx}. As first stressed by Truong, see
Ref.~\cite{Truong:1991gv} (and references therein), unitarization of
chiral scattering amplitudes can generate resonances -- however, this 
extension of CHPT to higher energies comes of course with a price, as one
resums certain classes of diagrams and thus can not make the direct
connection to  QCD Green functions easily.


\section{Pion-nucleon scattering}
\label{sec:piN}

\begin{figure}[t]
\begin{center}
\includegraphics[width=7cm,angle=0]{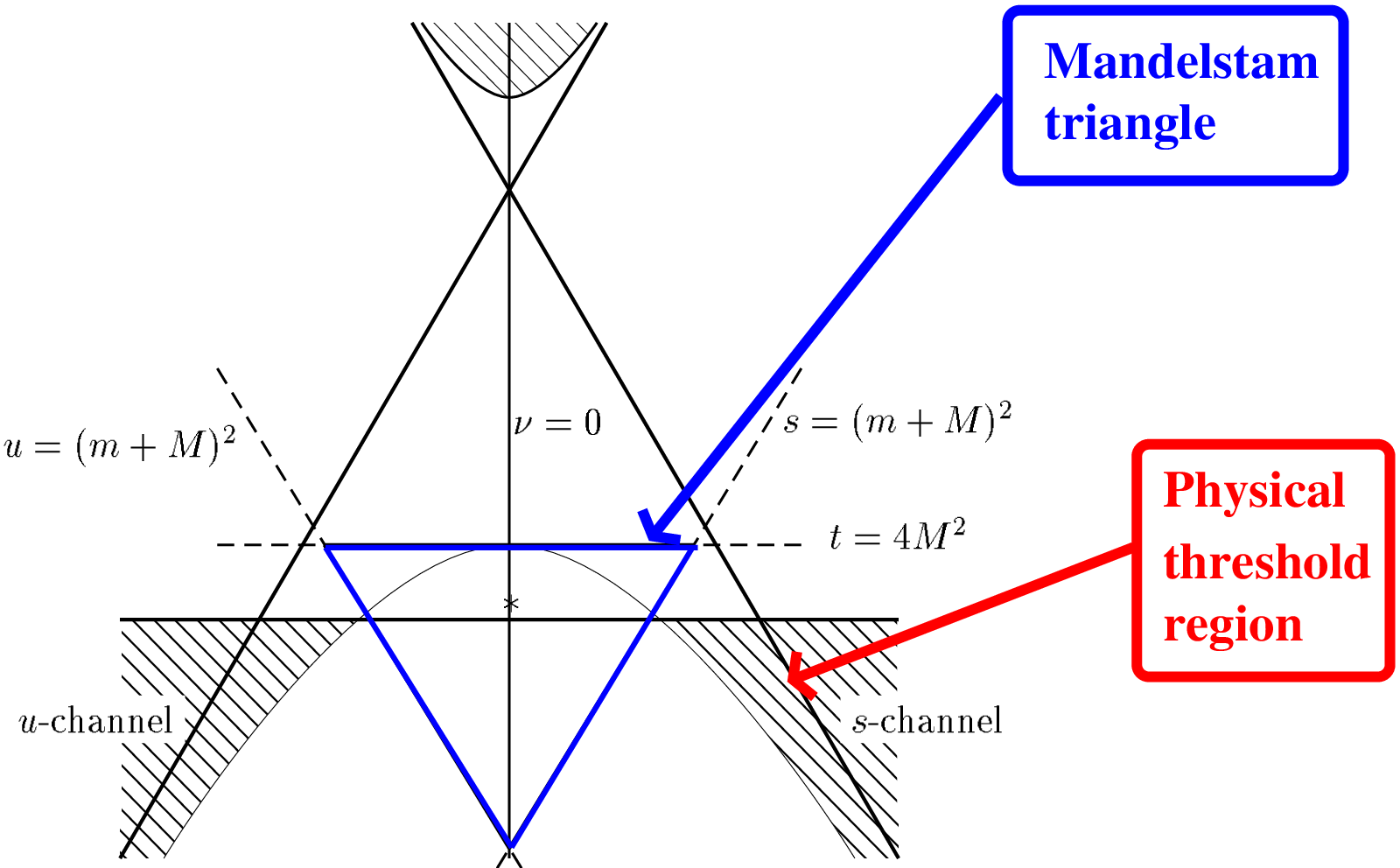}
\end{center}
\caption{\label{fig:mandel}
Mandelstam plane. The Mandelstam triangle and the threshold region 
for elastic scattering are indicated.}
\end{figure}

Pion-nucleon scattering is one of the premier reactions to test the chiral dynamics of QCD. It is also an
important ingredient in the description of the forces between two nucleons, as mentioned above
and will be made more explicit in a later section. The reaction $\pi^a (q)+N(p) \to \pi^b(q')+ N(p')$ is best described in 
terms of the Mandelstam variables, with $s = (p+q)^2$, $t =  (q'-q)^2$ and $u= (p-q')^2$, subject to the constraint
$s+t+u = 2(M_\pi^2+m_N^2)$, and $a,b$ are isospin indices. $M_\pi (m_N)$ is the pion (nucleon) mass.
The Mandelstam plane for this process is shown in 
Fig.~\ref{fig:mandel}. The threshold region for the $s$-channel process is the hatched area on the right side, 
the interior Mandelstam triangle (subthreshold region), where the scattering amplitude is real, is shown by the
triangle. These are the regions where chiral perturbation theory has been applied to pin down the so important 
LECs $c_i$. A first series of works, employing the heavy baryon approach, was performed at J\"ulich around the
year 2000 \cite{Fettes:1998ud,Fettes:2000xg}, pioneering also the matching 
of chiral amplitudes in the subthreshold region to a dispersive representation
from the Karlsruhe-Helsinki group \cite{Buettiker:1999ap}. 
The resulting values for the $c_i$'s are listed in the first row of Tab.~\ref{tab:ci}.
The second row gives a more recent determination from the Bochum group \cite{Krebs:2012yv}, 
where the range is due to the fit to
various partial wave analyses, accounting also for the different counting of the nucleon mass in the chiral EFT for
nuclear forces (as explained below). One notices that the errors are sizeable and that there are systematic differences.
This can be partly traced back to the fact that the chiral representation of the $\pi N$ scattering amplitude does not provide
sufficient curvature in the subthreshold region as first pointed out by 
Becher and Leutwyler~\cite{Becher:2001hv} . 

\renewcommand{\arraystretch}{1.2}
\begin{table*}[t!]
\begin{center}
\begin{tabular}{|c||c|c|c|c|}
\hline 
Method & $c_1$ & $c_2$ & $c_3$ & $c_4$ \\
\hline
CHPT (phase  shifts + subthr.) & $-0.9^{+0.2}_{-0.5}$ & $3.3 \pm 0.2$ & $-4.7^{+1.2}_{-1.0}$ & $3.5^{+0.5}_{-0.2}$\\
CHPT (phase shifts)                   &  $-1.13 ... -0.75$ & $3.49 ... 3.69$ & $-5.51 ... -4.77$ &  $3.34 ... 3.71$ \\
\hline
Roy-Steiner (standard counting)  & $ -1.11(3)$ & $3.13(3)$ & $-5.16(6)$ & $4.26(4)$ \\
Roy-Steiner (NN counting)            & $ -1.10(3)$ & $3.57(4)$ & $-5.54(6)$ & $4.17(4)$ \\
\hline
\end{tabular}
\caption{Extraction of some dimension two pion-nucleon LECs using CHPT and using Roy-Steiner equations. 
For details, see the text.
\label{tab:ci}}
\end{center}
\end{table*}

Before coming back to the determination of the LECs $c_i$, another important development deserves to be mentioned.
Triggered by extremely accurate measurements of the energy level shifts and widths of pionic hydrogen and deuterium at PSI~\cite{Strauch:2010vu,Hennebach:2014lsa},
the authors of Refs.~\cite{Baru:2010xn,Baru:2011bw} 
used CHPT to calculate the $\pi^- d$ scattering length, where $d$ denotes the deuteron,  with an accuracy 
of a few percent. In particular, for the first time isospin-violating corrections in the two- and three-body systems were included consistently. 
Using the PSI data on pionic deuterium and pionic hydrogen atoms, the isoscalar and isovector pion-nucleon scattering lengths
could be extracted with high precision,
\beqa\label{eq:a}
 a^+ &=& (7.6 \pm 3.1) \times 10^{-3}  M_{\pi^+}^{-1}~,\nonumber\\ 
 a^-  &=& (86.1 \pm 0.9) \times 10^{-3} M_{\pi^+}^{-1}~. 
 \eeqa
 This is truly a remarkable achievement. The famous lowest order predictions are $a^+ = 0$ and 
 $a^- = 79.4  \times 10^{-3} M_{\pi^+}^{-1}$~ \cite{Weinberg:1966kf,Tomozawa:1966jm}.
 For the first time, the sign of the small isoscalar scattering length could be fixed with $2.5 \, \sigma$ certainty.  Using  
 the Goldberger-Miyazawa-Oehme sum rule~\cite{Goldberger:1955zza}, 
 this leads to the  charged-pion-nucleon  coupling constant $g_c^2/(4 \pi) = 13.69 \pm 0.20$.  

\begin{figure}
\centering
\includegraphics[height=\linewidth,angle=-90]{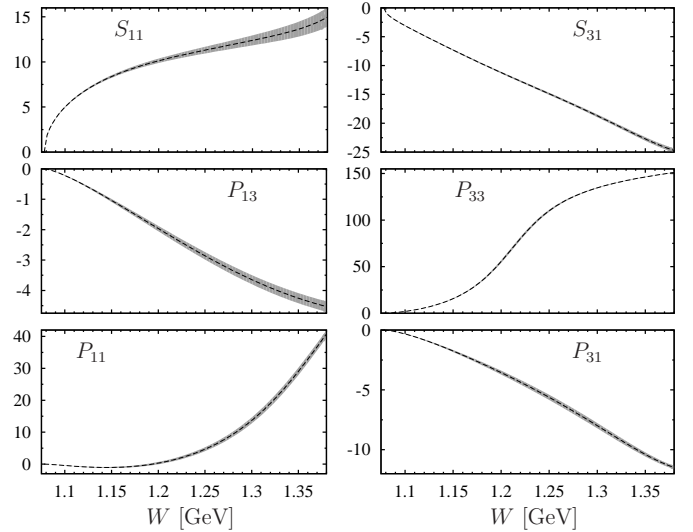}
\caption{Phase shifts $\delta_{l\pm}^{I_s}$ of the $s$-channel partial waves in degrees, obtained from the solution of the RS equations~\cite{Hoferichter:2015dsa}. 
The dashed line indicates the central solution, the bands the uncertainty estimate.
The partial waves are labeled by the spectroscopic notation $L_{2I_s2J}$.}
\label{fig:schannel}
\end{figure}

I return to the issue of determining the pion-nucleon scattering amplitude,
which is best done using dispersion relations. Such a method can be applied
to investigate  various
scattering processes, like $\pi\pi$, $\pi K$ or $\pi N$ scattering.
Roy equations~\cite{Roy:1971tc,Ananthanarayan:2000ht} for $\pi\pi$ scattering, or RS 
equations~\cite{Hite:1973pm,Buettiker:2003pp,Hoferichter:2011wk,Ditsche:2012fv} 
for non-totally-crossing-symmetric processes, incorporate the constraints 
from analyticity, unitarity, and crossing symmetry in the form of dispersion 
relations for the partial waves. They can be shown to be rigorously valid 
in a certain kinematic region, in the case of $\pi N$ scattering the upper limit is 
$\sm=(1.38\GeV)^2$~\cite{Ditsche:2012fv}. The integral contributions above 
$\sm$ as well as partial waves with $l>l_\text{m}$, with $l_\text{m}$ the 
maximal angular momentum explicitly included in the calculation, are 
collected in the so-called driving terms, which need to be estimated from 
existing PWAs, as do inelastic contributions below $\sm$. The free parameters 
of the approach are subtraction constants, which, in the case of $\pi\pi$ scattering, 
can be directly identified with the scattering lengths~\cite{Ananthanarayan:2000ht}, 
while for the solution of the $\pi N$ system it is more convenient to relate 
them to subthreshold parameters instead. The resulting system of coupled integral 
equations corresponds to a self-consistency condition for the low-energy phase 
shifts, whose mathematical properties were investigated in detail 
in Ref.~\cite{Gasser:1999hz}. Following~\cite{Ananthanarayan:2000ht}, 
the authors of Ref.~\cite{Hoferichter:2015dsa}
pursued the following solution strategy: the phase shifts are parameterized 
in a convenient way with a few parameters each, which are matched to input 
partial waves above $\sm$ in a smooth way.
To measure the degree to which the RS are fulfilled, a $\chi^2$-like function is defined according to
\beq
\label{chisqr}
\chi^2=\sum_{l,I_s,\pm}\sum_{j=1}^N\Bigg(\frac{\Re f_{l\pm}^{I_s}(W_j)-F\big[f_{l\pm}^{I_s}\big](W_j)}{\Re f_{l\pm}^{I_s}(W_j)}\Bigg)^2,
\eeq
where $\{W_j\}$ denotes a set of points between threshold and $\sqrt{\sm}$, $f_{l\pm}^{I_s}$ are the $s$-channel partial waves with isospin $I_s$, orbital angular momentum $l$, and total angular momentum $j=l\pm 1/2\equiv l\pm$, and $F\big[f_{l\pm}^{I_s}\big]$ the right-hand side of the RS equations.
In  Ref.~\cite{Hoferichter:2015dsa} $l_\text{m}=1$, $N=25$ (distributed equidistantly) are taken, and the number of subtraction constants are chosen
in such a way as to match the number of degrees of freedom predicted by the mathematical properties of the Roy equations~\cite{Gasser:1999hz}. It should be stressed that the form of the RS equations only reduces to that of Roy equations once the $t$-channel is solved, see~\cite{Ditsche:2012fv}. In the solution of the RS equations we minimize Eq.~\eqref{chisqr} with respect to the subtraction constants (identified with subthreshold parameters) and the parameters describing the low-energy phase shifts, while imposing~Eq.~\eqref{eq:a} as additional constraints.
The solution for the $s$-channel partial waves, expressed in terms of the phase shifts and including uncertainty estimates from these systematic studies as well as the uncertainties in the scattering lengths and the coupling constant, is shown in Fig.~\ref{fig:schannel}.
Apart from low-energy phase shifts, the RS solution provides a consistent set of subthreshold parameters. In particular, this allows to pin down the much discussed
pion-nucleon $\sigma$-term:
\beq
\sigma_{\pi N}=(59.1\pm 3.5)\MeV.
\eeq
A crucial ingredient in this determination are the precise scattering lengths
given in Eq.~(\ref{eq:a}). By combining this information
with the constraints from RS equations, the $\sigma$-term can be determined to a remarkable accuracy. From matching the results for the subthreshold parameters of pion-nucleon 
scattering obtained from a solution of Roy-Steiner equations
to chiral perturbation theory up to next-to-next-to-next-to-leading order, 
one can  extract the pertinent low-energy constants, 
including a comprehensive analysis of systematic uncertainties and correlations, as
shown in the third row of Tab.~\ref{tab:ci} \cite{Hoferichter:2015tha}.
Results for the LECs are also presented in the counting scheme 
usually applied in chiral nuclear effective field theory,
$\{p,\mpi\}/\mN=\Order(p^2)$~\cite{Weinberg:1991um} (see below), are
shown in the fourth row of Tab.~\ref{tab:ci}. One sees that these parameters
are now determined with much better precision than before.


\section{Nuclear forces}
\label{sec:forces}

In this section, I discuss the construction and status of the nuclear forces derived in chiral EFT.
This is based on the ground-breaking work by Weinberg~\cite{Weinberg:1990rz,Weinberg:1991um,Weinberg:1992yk}
and by van Kolck~\cite{vanKolck:1994yi}. There has been much discussion about the Weinberg power counting, see e.g.
the review~\cite{Epelbaum:2008ga} or the recent talk by Phillips~\cite{Phillips:2013fia}. Space forbids to go into these
details, I will rather stick to this framework, which has proven to be extremely successful and discuss its
foundations and some very recent developments that not only allow for very precise calculations
of the forces between nucleons but also supply serious uncertainty estimates, going beyond the cut-off variations
mostly employed before.

\begin{figure}[ht]
\begin{center}
\includegraphics[width=0.450\textwidth,angle=0]{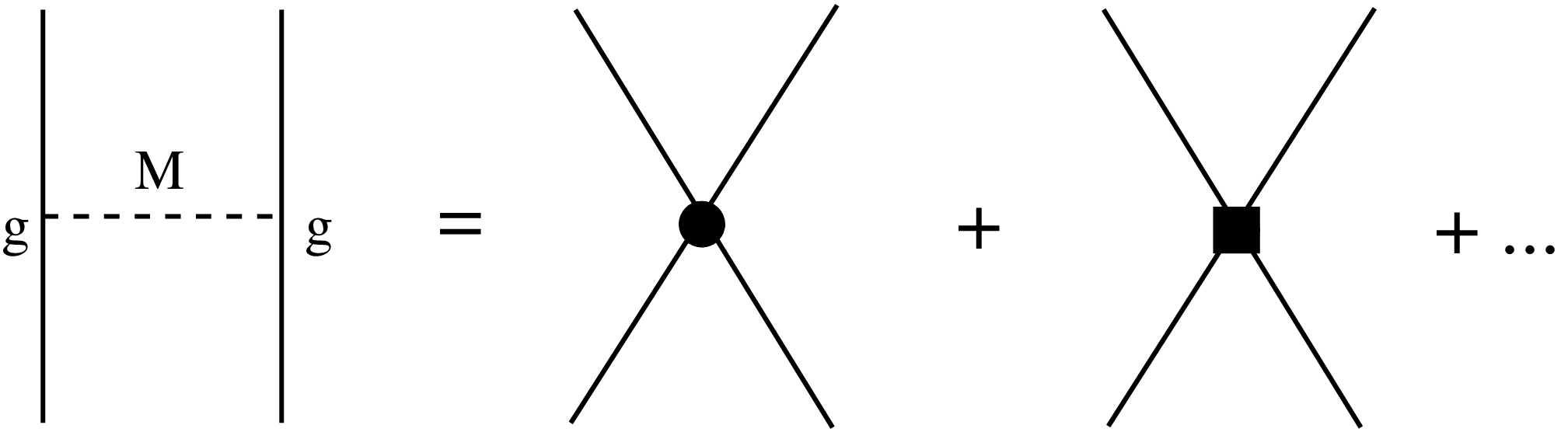}
\end{center}
\caption{\label{fig:resosat}
Resonance saturation for the dimension zero (circles) and two (squares)
LECs. In the limit that the meson resonance masses $M$ go to infinity,
but keeping $g^2/M$ fixed, with $g$ the meson-nucleon coupling constant,
one obtains a series of contact interactions with increasing number of derivatives.
}
\end{figure}

For  developing a systematic and model-independent theoretical framework capable to describe 
reactions involving several nucleons up to center-of-mass three-momenta of (at
least) the  order of the pion mass $M_\pi$, one has to realize that
nuclear binding is very shallow,  with typical binding energies per nucleon much smaller
than the pion mass. The appearance of shallow bound states can not be described in 
perturbation theory. The quest for an EFT  that allows for the  most general parameterization of 
the nucleon-nucleon scattering  amplitude consistent with the fundamental principles 
such as Lorentz invariance, cluster separability and analyticity, must account for this and other  basic facts of 
nuclear physics. The energies of the nucleons we are interested in are well below the nucleon mass, it therefore is
natural and  appropriate to make use of  a non-relativistic expansion, that is an expansion in inverse powers of the 
nucleon mass $m_N$.  Accordingly, in the absence of external probes and below the pion production 
threshold, one is left with a potential theory in the framework of 
the quantum-mechanical $A$-body Schr\"odinger equation
\beq
\label{SE}
\big( H_0  + V \big) |
\Psi \rangle = E | \Psi \rangle\,,  ~~
\mbox{with} 
~~ H_0 = \sum_{i=1}^A \frac{-\vec \nabla_i^2}{2
  m_N}  + \mathcal O (m_N^{-3})\,.
\eeq
The main task then reduces to the determination of the nuclear
Hamilton operator $H_0 +  V$. This can be 
accomplished using the framework of CHPT. 
It is further important to realize that the nuclear interactions feature two very distinct
contributions, long-range one- and two-pion exchanges and shorter-ranged
interactions, that can be represented by a tower of multi-nucleon operators.
As the pion is the pseudo-Goldstone boson of the approximate chiral symmetry
of QCD as discussed above, its interactions with  the 
nucleons are of derivative nature and strongly constrained by the available
data on pion-nucleon scattering and other fundamental processes. However,
in harmony with the principles underlying EFT, one has also to consider operators of nucleon fields only. 
In a meson-exchange model of the nuclear forces, these can be pictured 
by the exchanges of heavier mesons like $\sigma$, $\rho$, $\omega$, and 
so on, see Fig.~\ref{fig:resosat} -- but such a modeling is no longer necessary and also does not 
automatically generate all structures consistent with the underlying
symmetries. Also, in the EFT approach, the forces between three and four
nucleons are generated consistently with the dominant two-nucleon forces -
which could never be achieved in earlier modeling of these forces.

Within the framework of CHPT, nuclear forces are 
derived from the most general effective chiral Lagrangian by making
an expansion in powers of the small parameter $q$ defined
as
\beq
\label{deltaless}
q \in \bigg\{ \frac{M_\pi}{\Lambda}, \,    \frac{| \vec k\, | }{\Lambda} \bigg\}\,,
\eeq
where $Q \sim | \vec k | \sim M_\pi$ is a typical external momentum (the soft
scale)\footnote{We use the small parameter $q$ and the soft scale $Q$ synonymously.} 
and $\Lambda$  is a hard scale, sometimes also called breakdown scale. 
Appropriate powers of the inverse of this scale determine the
size of  the renormalized LECs in the effective Lagrangian.  
Notice that once renormalization of loop contributions is carried out and the renormalization scale 
is set to be $\mu \sim M_\pi$ as appropriate in CHPT, all momenta flowing
through diagrams appear to be, effectively, of the order $ \sim M_\pi$.
Consequently, one can use naive dimensional analysis 
to estimate the importance of (renormalized) contributions of
individual diagrams. 

\begin{figure*}[t!]
\begin{center}
\includegraphics[width=14cm,angle=0,clip]{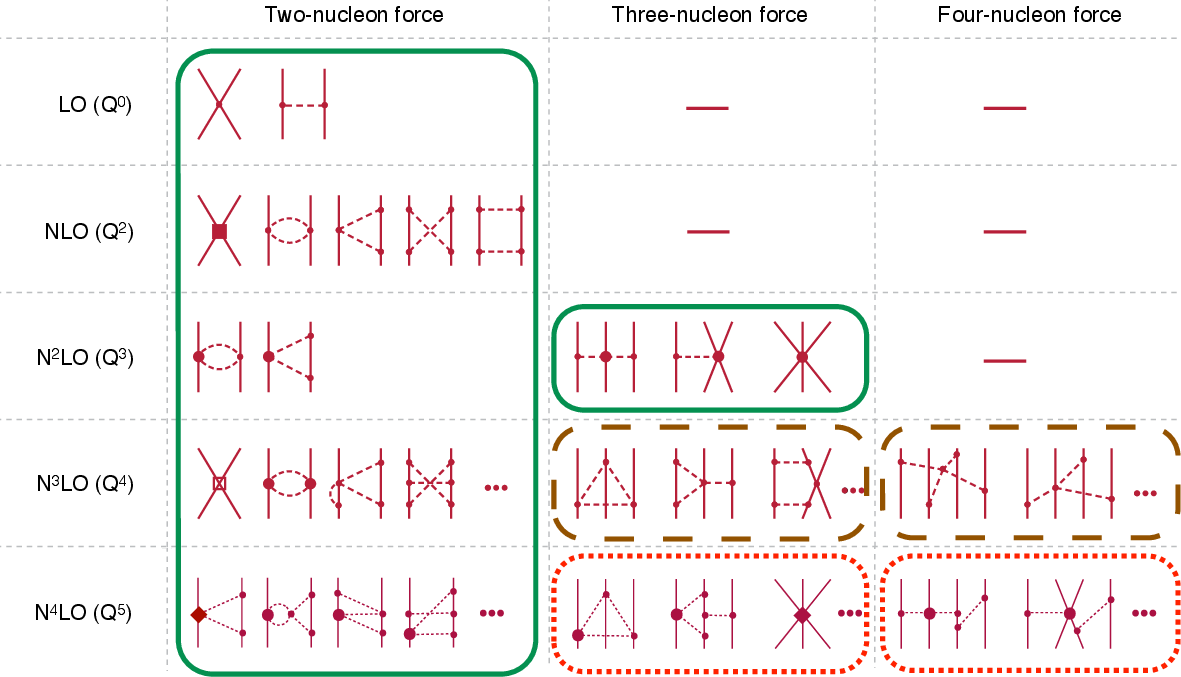}
\end{center}
\caption{\label{fig:forces}
Contributions to the effective potential of the 2N, 3N and 4N forces
based on Weinberg's power counting.
Here, LO denotes leading order, NLO next-to-leading order and so on.
The various vertices according to Eq.~(\ref{powc_w}) with $\Delta_i = 0,1,2,3,4$
are denoted by small circles, big circles,  filled boxes, filled diamonds and open boxes, 
respectively.  The boxes surrounding various classes of diagrams are explained in the text.
Figure courtesy of Evgeny Epelbaum.
}
\end{figure*}

To be specific, consider a connected 
Feynman graph with $N$ nucleon lines.\footnote{Note that nucleons cannot be destroyed 
or created within the non-relativistic approach.} It is easier to count the powers 
of the hard scale $\Lambda$ rather than of the soft scale $Q$ by observing 
that the only way for $\Lambda$
to emerge is through the corresponding LECs, as first pointed out by Epelbaum~\cite{Epelbaum:2006eu,Epelbaum:2007us}.
Thus, the  low-momentum dimension $\nu$ of a given diagram can be expressed in terms of the canonical 
field dimensions $\kappa_i+4$  of $V_i$ vertices of type $i$  via
\beq
\label{pow_co_alg}
\nu = -2 + \sum V_i \kappa_i \,, \quad \quad \kappa_i = d_i + \frac{3}{2} n_i + p_i -4 \,,
\eeq
where $n_i$ ($p_i$) and $d_i$ refer to the number of the nucleon (pion) field operators and derivatives or 
pion mass insertions, respectively. The constant $-2$ in the expression for
$\nu$ is just a convention.   
The power counting can also be re-written in terms of topological variables such as the number of 
loops $L$ and nucleon lines $N$ rather than $\kappa_i$ which are appropriate for diagrammatic approaches. 
 For connected diagrams the above equation then takes the form  
\beq
\label{powc_w}
\nu = -4 + 2N + 2 L + \sum V_i \Delta_i \,, ~~~ \Delta_i = d_i + \frac{1}{2} n_i -2 \,.
\eeq
Chiral symmetry
of QCD guarantees that the pions couple only through vertices
involving derivatives or powers of $M_\pi$.  This implies 
that the effective Lagrangian contains only irrelevant (i.e.~non-renormalizable) interactions with 
$\kappa_i \geq 1$ ($\Delta_i \geq 0$) which allows for a
\emph{perturbative} description of pion-pion and pion-nucleon scattering
as well as nuclear forces. 
The leading interactions, i.e.~the ones  with the smallest possible
$\Delta_i$, that is $\Delta_i=0$,  have the form 
\beqa
\label{lagr}
 \mathcal{L}^{(0)} \!\!\!\!\! &=& \!\!\!\!\! \frac{1}{2} \partial_\mu \fet \pi \cdot \partial^\mu \fet \pi - 
\frac{1}{2}  M_\pi^2 \fet \pi^2 \nn 
&+& \!\!\!\!\!\!  N^\dagger \left[ i \partial_0 + \frac{g_A}{2 F_\pi} \fet \tau \vec \sigma 
\cdot \vec \nabla \fet \pi - \frac{1}{4 F_\pi^2} \fet \tau \cdot (\fet \pi \times \dot{\fet \pi}) \right] N\nn
&-& \!\!\!\!\! \frac{1}{2} C_S (N^\dagger N) (N^\dagger N) -  \frac{1}{2} C_T (N^\dagger \vec \sigma N) \cdot (N^\dagger 
\vec \sigma N) + \ldots \,, \nn &&
\eeqa
where $\fet \pi$ and $N$ refer to the pion and nucleon field operators, respectively, 
and $\vec \sigma$ ($\fet \tau$) denote the spin (isospin) Pauli matrices.  Further, $g_A$ ($F_\pi$) is the nucleon 
axial-vector coupling (pion decay) constant and $C_{S,T}$ are the LECs
accompanying the leading contact  operators. 
The ellipses refer to terms involving more pion fields. It is
important to emphasize that chiral symmetry leads to highly nontrivial 
relations between the various coupling constants. For example, the strengths of all $\Delta_i=0$-vertices without 
nucleons with $2, 4, 6, \ldots$ pion field operators are given in terms of $F_\pi$ and $M_\pi$. Similarly, 
all single-nucleon $\Delta_i=0$-vertices with $1,2,3,\ldots$ pion fields are expressed in terms of just 
two LECs, namely  $g_A$ and  $F_\pi$.  
The construction of the higher order terms is well documented in the 
literature~\cite{Bernard:1995dp,Epelbaum:2008ga,Fettes:2000gb,Machleidt:2011zz,Scherer:2012xha}.

The expressions for the power counting given above are derived under the assumption that there are no
infrared divergences. This assumption is violated for a certain class of diagrams involving two and more nucleons 
(more precisely, for the two-nucleon case the so-called box diagram is the culprit) due to the
appearance of  pinch singularities of the kind~\cite{Weinberg:1991um} 
\beq
\label{example}
\int d l_0 \frac{i}{l_0 + i \epsilon} \,  \frac{i}{l_0 - i \epsilon}
\,.
\eeq
Here, $i/(l_0 + i \epsilon)$ is the free nucleon propagator in the
heavy-baryon approach (in the nucleon rest-frame) corresponding to the Lagrangian in
Eq.~(\ref{lagr}). 
Clearly, the divergence
is not ``real'' but just an artifact of the extreme non-relativistic approximation
for the propagator which is not applicable in that
case. Keeping the first correction beyond the static limit, the
nucleon propagator takes the form $i/(l_0 - \vec l \,^2/(2m_N) + i
\epsilon)^{-1}$ leading to a finite result for the integral in
Eq.~(\ref{example}) which is, however, enhanced by a factor $m_N / | \vec
q \, |$ as compared to the estimation based on naive dimensional
analysis. In physical terms, the origin of this enhancement is related
to the two-nucleon Green's function of the Schr\"odinger equation
(\ref{SE}). The nuclear potential $V$ we are actually interested in
is, of course,  well defined in the static limit $m_N \to \infty$ and thus
not affected by the above mentioned infrared enhancement.  More precisely,
the potential is defined in terms of the so-called irreducible contributions and
all reducible contributions that are generated from the
iteration of the potential in the Schr\"odinger equation.

It is now instructive to address the qualitative implications of the power
counting in Eq.~(\ref{powc_w}) and the explicit form of the effective
chiral Lagrangian.  First, one observes that the dominant contribution
to the nuclear force arises from two-nucleon tree-level diagrams with
the lowest-order
vertices. This implies that the nuclear force is dominated by the
one-pion exchange potential and the two contact
interactions without derivatives. Pion loops are suppressed by two powers of the soft
scale. Also vertices with $\Delta_i >0 $  involving more derivatives   
are suppressed and do not contribute at lowest order. One also
observes the suppression of many-body forces: according to
Eq.~(\ref{powc_w}), $N$-nucleon forces start contributing at order 
$Q^{-4 + 2N}$.  This implies the dominance of the two-nucleon force
with three- and four-nucleon forces  appearing formally  as corrections at
orders $Q^2$ and $Q^4$, respectively.  However, as pointed out by Weinberg and 
van Kolck, the leading irreducible contributions to the  three-nucleon potential cancel,
so that three-nucleon forces indeed start at order $Q^3$. All this is
summarized in Fig.~\ref{fig:forces}. The present state-of-the-art of deriving the 
nuclear Hamiltonian is also shown in this figure.  The green (solid) boxes show the 
parts of the potential that have been worked out and applied, the brown (long-dashed) ones
refer to contributions that are also available but are only being
included in explicit calculations of observables now and the red (short-dashed) 
boxes refer to contributions still in the process of
being worked out. As can be seen from this, calculations within the
two-nucleon system are by far most advanced, and I therefore will describe the
most recent developments here. First, it is important to note that in 
Ref.~\cite{Epelbaum:2014efa} a new coordinate space regularization was
introduced (see also Refs.~\cite{Rentmeester:1999vw,Gezerlis:2013ipa,Gezerlis:2014zia}), that 
does not lead to any  distortion of the long-range part of the potential as
the earlier used momentum cut-off:
\beq
V_{\rm long-range} ( \vec r \, )  \to V_{\rm long-range}^{\rm reg} (
\vec r \, ) = V_{\rm long-range} ( \vec r \, ) f \left( \frac{r}{R}
\right)\,,
\eeq
where the regulator function $f(x)$ is chosen such that its value  goes to 
$0$ $(1)$ sufficiently fast for $x \to 0$ (exponentially fast for $x \gg 1$).
A further advantage of this scheme is that the above choice of the
regulator makes the additional spectral function regularization  of the pion
exchange contributions obsolete. The regulator function $f(r/R)$  can be chosen 
as
\beq
\label{NewReg}
f \left( \frac{r}{R} \right) = \left[1- \exp\left( - \frac{r^2}{R^2} \right)\right]^n\,,
\eeq
where the exponent $n$ has to be taken sufficiently large. It is
necessary to choose $n = 4$ or larger in order to make the regularized
expressions for the dimensionally regularized two-pion exchange potential 
at N$^3$LO vanish in the origin, however, larger values of $n$ lead to more stable
numerical results when doing calculations in momentum space. So $n=6$ was chosen in
Refs.~\cite{Epelbaum:2014efa,Epelbaum:2014sza}. In fact, in Ref.~\cite{Epelbaum:2014efa}
independence  of observables for $n \geq 5$ is explitely demonstrated.
Another important progress made
in Ref.~\cite{Epelbaum:2014efa} was the introduction of a better scheme to 
quantify the theoretical uncertainties. For that, one first has to analyze the
possible sources of uncertainties (see also Refs.~\cite{Furnstahl:2014xsa,Perez:2014kpa}). 
These include 1) the {\em systematic} uncertainty due to truncation of the chiral
expansion at a given order, 2) the uncertainty in the knowledge of $\pi N$ LECs 
which govern the long-range part of the nuclear force, 3) the
uncertainty in the determination of LECs accompanying the contact interactions; 
and 4) uncertainties in the experimental data or, in the partial wave analysis if that
is used to determine the LECs. As described above, there has been much progress
in determining the $\pi N$ LECs, so we concentrate on the first type of uncertainty.
For a given observable $X( p)$, where $p$ is the center-of-mass momentum
corresponding to the considered energy, the expansion parameter in
chiral EFT is given by Eq.(\ref{deltaless}),
where $\Lambda$ is the breakdown scale. As discussed in  Ref.~\cite{Epelbaum:2014efa},
one should use $\Lambda = 600\,$MeV for the cutoffs $R=0.8$, $0.9$ and
$1.0\,$fm, $\Lambda = 500\,$MeV for $R=1.1\,$fm and $\Lambda =400\,$MeV
for $R=1.2$ to account for the increasing amount of cutoff artifacts. In fact, 
when increasing the $r$-space cutoff $R$, one actually continuously 
integrates out pion physics, and the resulting theory would gradually turn
into pionless EFT if one would further soften the cutoff. Having verified
this estimation of the breakdown scale on the example of the  neutron-proton 
scattering total cross 
section at various chiral orders~\cite{Epelbaum:2014efa}, one is naturally led to 
a method that gives a conservative estimate of the theoretical uncertainty due to the neglect of higher orders.
In this approach, one ascribes the uncertainty $\Delta X^{\rm N^4LO} (p)$ of a N$^4$LO 
prediction $X^{\rm N^4LO}(p)$ for an observable $X(p)$, as (and similarly for lower orders)  
\begin{eqnarray}
\label{def_error}
\Delta X^{\rm N^4LO} (p) &=& \max  \bigg( Q^6 \times \Big| X^{\rm
    LO}(p) \Big|, \\
&& {}  \;\;\;\;\;\; \; \; \; \;  Q^4 \times \Big|
  X^{\rm LO}(p) -   X^{\rm NLO}(p) \Big|, \nn  
&&{}  \;\;\;\;\;\; \; \; \; \;  Q^3 \times \Big|
  X^{\rm NLO}(p) -   X^{\rm N^2LO}(p) \Big|, \nn
&& {}  \;\;\;\;\;\; \; \; \; \;  Q^2 \times \Big|
  X^{\rm N^2LO}(p) -   X^{\rm N^3LO}(p) \Big|  , \nn
&& {}  \;\;\;\;\;\; \; \; \; \;  Q \times \Big|
  X^{\rm N^3LO}(p) -   X^{\rm N^4LO}(p) \Big|  \bigg)\,,
\nonumber
\end{eqnarray}
where the expansion parameter $Q$ is given by Eq.~(\ref{deltaless}) and   
the scale $\Lambda$ is chosen dependent of the cutoff $R$ as
discussed above. The resulting theoretical uncertainties for the total cross section and 
the case of R = 0.9 fm  were found in Ref.~\cite{Furnstahl:2015rha}  to be consistent with 
the 68\% degree-of-belief intervals for EFT predictions.

\begin{figure}[b]
\begin{center}
\includegraphics[width=0.45\textwidth,keepaspectratio,angle=0,clip]{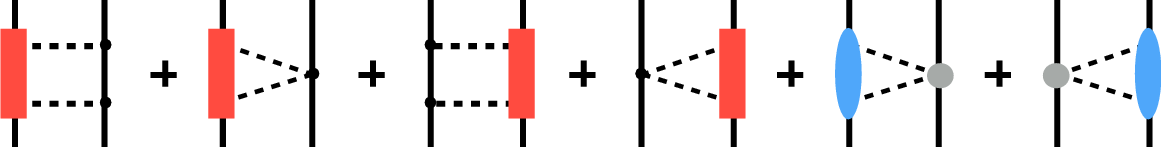}
\end{center}
    \caption{Fifth order contributions to the two-pion exchange potential. 
Solid and dashed lines refer to  nucleons and pions, respectively. 
Solid dots denote vertices from the lowest-order $\pi N$ effective Lagrangian. Filled 
rectangles,  ovals and grey circles  denote the order $Q^4$, order $Q^3$ and order $Q^2$ 
contributions to $\pi N$ scattering, respectively.   
\label{Fig:diagrams} 
 }
\end{figure}

\begin{figure}[t]
\includegraphics[width=0.49\textwidth,keepaspectratio,angle=0,clip]{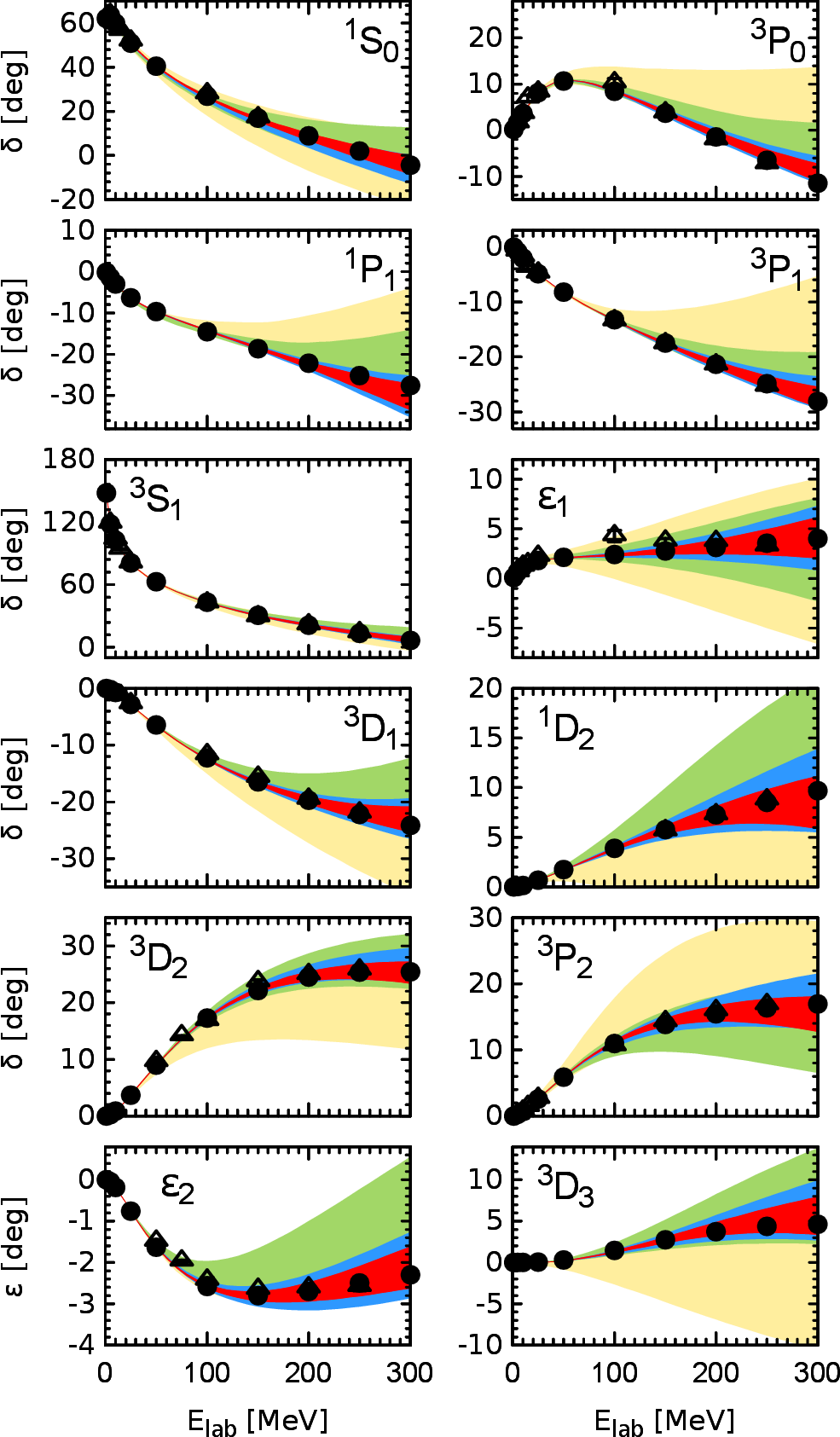}
\caption{Results for the np S-, P- and D-waves and the mixing angles $\epsilon_1$,  
$\epsilon_2$ up to N4LO  based on the cutoff of $R=0.9\,$fm in comparison with the 
Nimjegen PWA \cite{Stoks:1993tb}  and the GWU single-energy PWA \cite{Arndt:1994br}.
The bands of increasing width show estimated theoretical uncertainty at 
N$^4$LO, N$^3$LO, N$^2$LO and NLO.   
\label{fig:phases} 
 }
\end{figure}

The most sophisticated calculation in the two-nucleon
system is indeed the fifth order result by Epelbaum et al.~\cite{Epelbaum:2014sza},
which included {\em all} new two-pion exchange corrections appearing at this
order as shown in Fig.~\ref{Fig:diagrams} (see also the less complete work in
Refs.~\cite{Entem:2014msa,Entem:2015xwa}).
Although three-pion exchange formally appears at N$^3$LO and at N$^4$LO, it has usually been
neglected, as the (nominally) leading $3\pi$ exchange potential at N$^3$LO is
known to be weak compared to the two-pion exchange \cite{Kaiser:1999ff,Kaiser:1999jg} 
and to have negligibly small effect on phase shifts. However, the subleading 
corrections at N$^4$LO are enhanced due to the appearance of the LECs $c_i$  \cite{Kaiser:2001dm}.
To check the assertion that the $3\pi$ exchange can still be neglected, the authors
of Ref.~\cite{Epelbaum:2014sza} have carried out a N$^4$LO fit  
for the intermediate value of the cutoff of $R=1.0\;$fm, in 
which the dominant class-XIII $3\pi$  exchange potential $V_{3 \pi}^{\rm XIII}$ from
Ref.~\cite{Kaiser:2001dm} was  \emph{explicitly} included. No significant (not even
noticeable) changes 
both in the quality of the description of the Nijmegen phase shifts and 
in the reproduction/predictions for observables was found.
In Fig.~\ref{fig:phases}, using the above discussed method of uncertainty
quantification,  the S-, P- and D-wave phase shifts and the mixing
angles $\epsilon_1$ and $\epsilon_2$ at NLO and higher orders in the chiral expansion 
for $R=0.9\,$fm are shown. The various bands result from adding/subtracting the estimated
theoretical uncertainty  to/from the calculated results.
Similar results are obtained for np scattering observables, see Ref.~~\cite{Epelbaum:2014sza}
for details.

\begin{figure}[t]
\begin{center}
\includegraphics[width=5.0cm,angle=0]{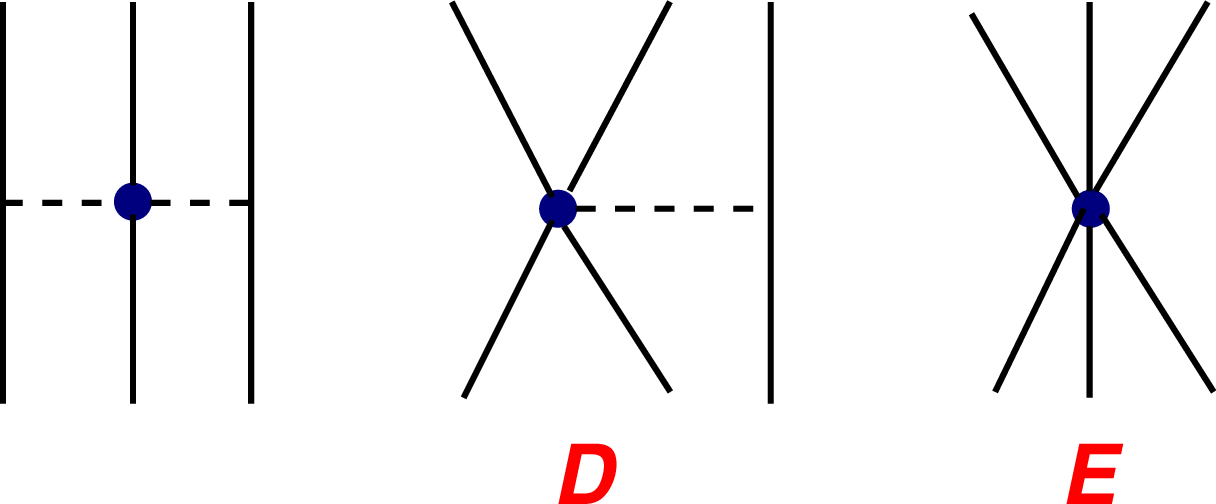}
\end{center}
\vspace{-3mm}
\caption{\label{fig:3nfdia}
Topologies of the leading contributions to the chiral 3NF. From left to
right: Two-pion exchange, one-pion-exchange and 6N contact interaction.
}
\end{figure}

\begin{figure}[b]
\begin{center}
\includegraphics[width=0.48\textwidth,angle=0]{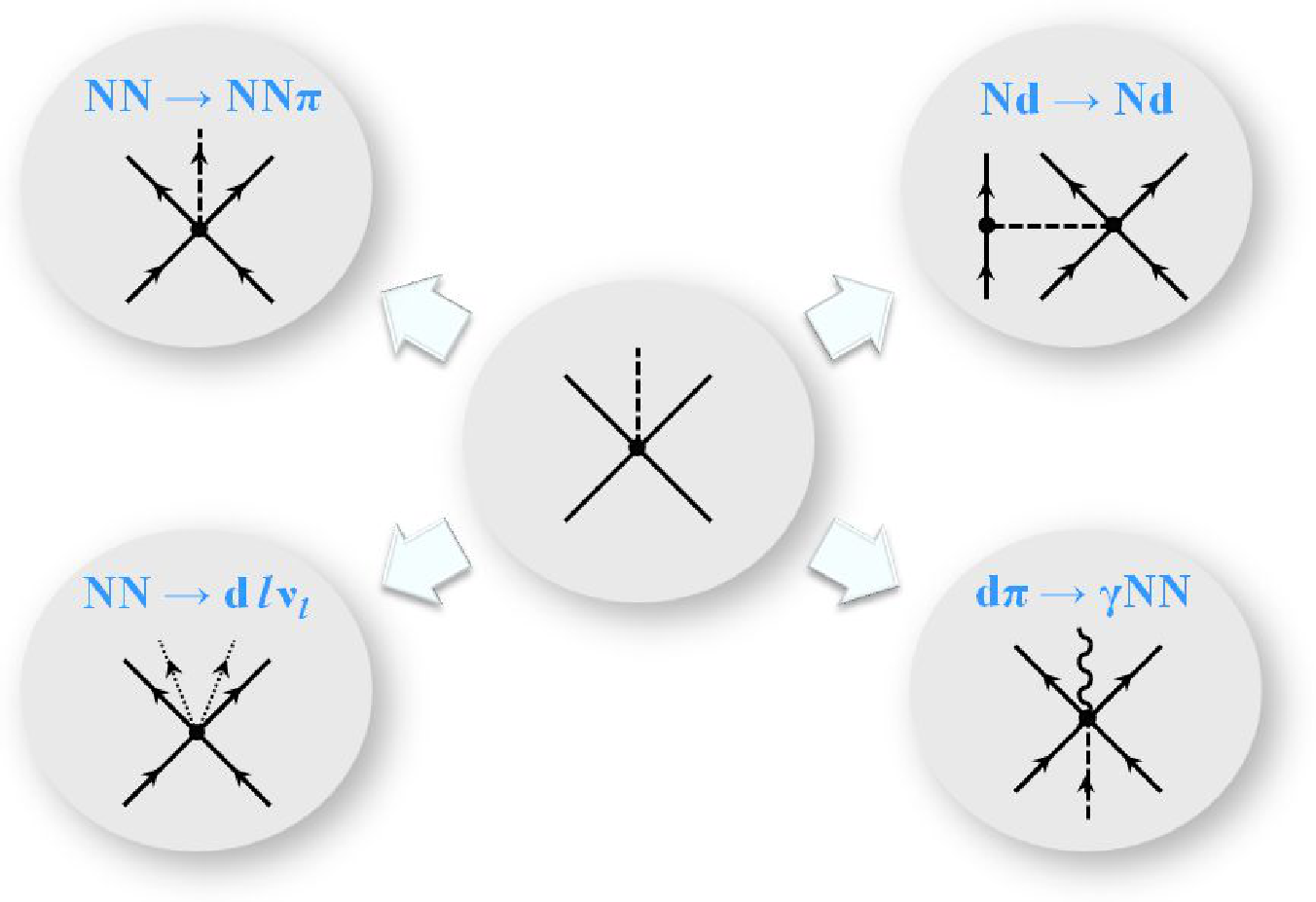}
\end{center}
\vspace{-3mm}
\caption{\label{fig:D}
Various reactions that all are sensitive to the $D$-term. Figure courtesy of
Evgeny Epelbaum.
}
\end{figure}

Next, let us consider three-nucleon forces (3NFs). While providing a small correction to the nuclear
Hamiltonian as compared to the dominant NN force, its inclusion is
mandatory for quantitative understanding of nuclear structure and 
reactions, for recent reviews, see Refs.~\cite{KalantarNayestanaki:2011wz,Hammer:2012id}.  
Historically, the importance of the 3NF has been pointed out
already in the 1930ties \cite{Primakoff:1939zz} while the first phenomenological 
3NF models date back to the 1950ties. However, in spite of
extensive efforts, the spin structure of the 3NF is still poorly
understood \cite{KalantarNayestanaki:2011wz}.
Chiral EFT indeed provides a suitable theoretical resolution to the
long-standing 3NF problem.  As already noted,
the three-nucleon force (3NF) only appears two orders after the leading NN 
interaction. At this order, there are only three topologies contributing,
see Fig.~\ref{fig:3nfdia}.
\noindent The two-pion exchange topology is given again in terms of the $c_i$,
as  discussed in detail in~\cite{Friar:1998zt}. The so-called $D$-term, which 
is related to the one-pion exchange between a 4N contact term and a further
nucleon, has gained some prominence in the first decade of this millennium,
as many authors have
tried to pin it down based on a cornucopia of reactions, 
such as $Nd \to Nd$~\cite{Epelbaum:2002vt},
$NN \to NN\pi$~\cite{Hanhart:2000gp,Baru:2009fm}, 
$NN\to d \ell \nu_\ell$~\cite{Ando:2002pv,Park:2002yp,Nakamura:2007vi,Gazit:2008ma},
$d \pi \to \gamma NN$~\cite{Lensky:2005hb,Gardestig:2005pp,Lensky:2007zc}, 
or the spectra of light nuclei~\cite{Nogga:2005hp},
see Fig.~\ref{fig:D} (here, $\gamma$ denotes a photon, $\ell$ a lepton and $\nu_\ell$ its corresponding antineutrino).
This demonstrates again the power of EFT -
very different processes are related through the same LECs thus providing many
different tests of chiral symmetry (as it is also the case with the
LECs $c_i$, see Fig.~\ref{fig:ci}). The LEC $E$ related to the 6N contact 
interaction can only be fixed in systems with at least three nucleons, say
from the triton binding energy. Indeed, the leading
chiral 3NF has already been extensively explored in \emph{ab initio} calculations by various
groups and found to yield promising results for nuclear structure
and reactions \cite{Hammer:2012id,Barrett:2013nh}.  
The first corrections to the 3NF at order $Q^4$ (N$^3$LO)
have also been derived \cite{Ishikawa:2007zz,Bernard:2007sp,Bernard:2011zr} 
(and are parameter-free) while the sub-subleading contributions at order $Q^5$ (N$^4$LO)
are being derived \cite{Girlanda:2011fh,Krebs:2012yv,Krebs:2013kha}.  The LENPIC
collaboration\footnote{LENPIC stands for Low Energy Nuclear Physics International Collaboration.}
\cite{lenpic} aims at working out the consequences of the sub- and sub-sub-leading corrections
to the 3NFs in light and medium nuclei. As a first step, utilizing the
fifth order two-nucleon forces and the method for error quantification discussed before, 
LENPIC studied nucleon-deuteron (Nd) scattering and selected low-energy observables in $^3$H, 
$^4$He, and $^6$Li based on NN forces only. 
Calculations beyond second order differ from experiment well outside the 
range of the quantified uncertainties~\cite{Binder:2015mbz}. This provides truly 
unambiguous evidence for missing
three-nucleon forces within the employed framework.
\begin{figure}[h]
\includegraphics[width=0.49\textwidth,keepaspectratio,angle=0,clip]{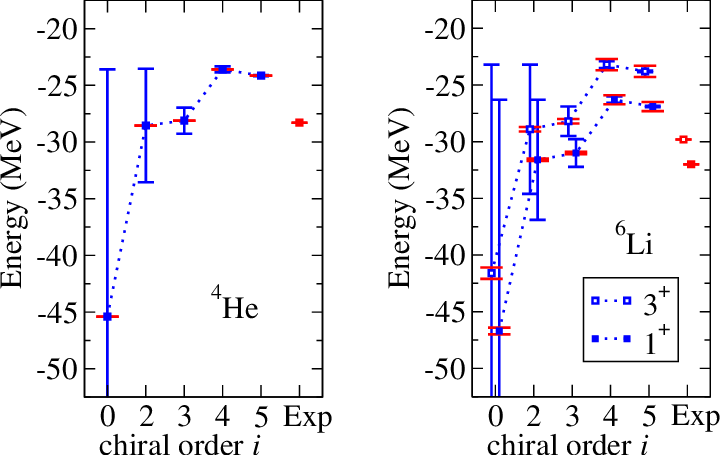}
\caption{Predictions for $E_{\rm gs}$ of $^4$He 
    and the energies of the lowest two states of $^6$Li based on the
  NN potentials of Refs.~\cite{Epelbaum:2014efa,Epelbaum:2014sza} for $R=1.0$~fm without 
  including the 3NF. Theoretical uncertainties are given by solid narow lines (blue) as detailed
  in Ref.~\cite{Binder:2015mbz}.  Numerical uncertainties from the NCSM solid wide lines (red) 
  are estimated following Ref.~\cite{Maris:2013poa}.
\label{fig:he4li6}}
\end{figure}

Four-nucleon forces (4NFs)  appear first at N$^3$LO and have been worked out
some time ago~\cite{Epelbaum:2006eu,Epelbaum:2007us}.
A rough estimate of its contribution to the $^4$He binding energy was performed in Ref.~\cite{Rozpedzik:2006yi}.
It was shown  that the four-nucleon force is attractive for wave functions
with a totally symmetric momentum part and the additional binding energy provided by the long-ranged part of the 4NF 
is of the order of a few hundred keV. However, in heavier nuclei, the four-nucleon forces must play a more important role, 
but explicit calculations need to be performed. Pioneering calculations
exploring the role of chiral 4NFs in nuclear matter have been performed in 
Refs.~\cite{Kaiser:2012ex,Kaiser:2015lsa}. Strong cancellations are found between various
types of contributions, but still an attractive and non-negligible contribution to the
binding energy of nuclear matter at saturation density is found. The role of 4NFs in the neutron 
matter equation of state was investigated in Ref.~\cite{Kruger:2013kua}.


\section{Discretization of space-time}
\label{sec:latt}

There are two different venues to tackle
the nuclear many-body problem, that is nuclei with atomic number $A\geq 5$.
Either one utilizes the forces from EFT within a conventional, well established
many-body technique (no-core-shell-model, coupled cluster approach, etc.)
or one develops a novel scheme that combines these forces with Monte Carlo
methods that are so successfully used in lattice QCD. This new scheme is
termed ``nuclear lattice simulations'' or ``Nuclear Lattice Effective Field Theory''
(NLEFT)  and has  enjoyed  wide recognition in the popular press as
the first ever {\sl ab initio} calculation of the Hoyle state in
$^{12}$C has been performed, see Sec.~\ref{sec:res}. In the following, 
I will give a short introduction
into this novel nuclear many-body technique. The foundations of the method and its
early applications are reviewed in Ref.~\cite{Lee:2008fa}.

\begin{figure}[!t]
  \centering
  \includegraphics[width=0.48\textwidth]{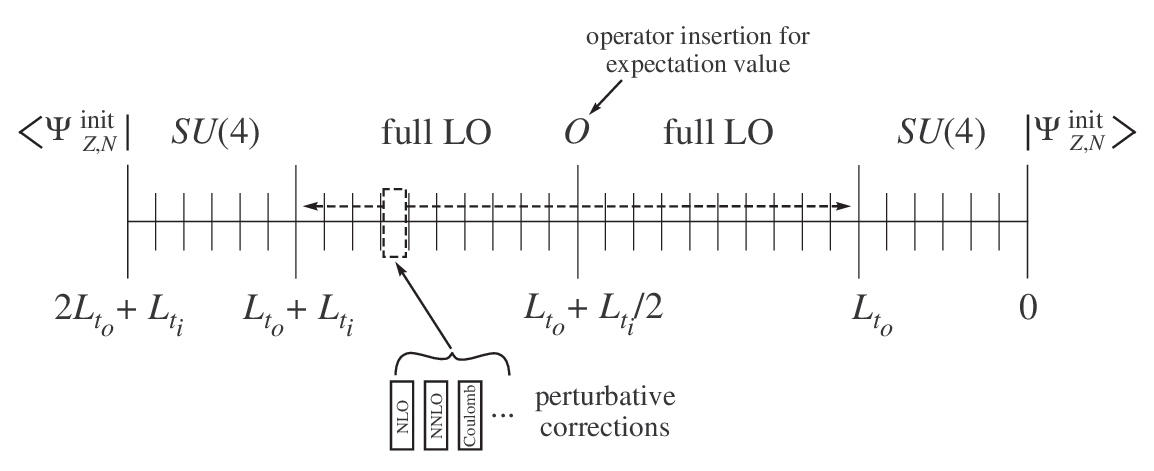}
  \caption{\footnotesize 
  Schematic diagram for the transfer matrix calculation. For details, 
  see the text.}
  \label{fig:time}
\end{figure}

Space-time is discretized in Euclidean time on a torus of volume
$L_s\times L_s\times L_s\times L_t$, with $L_s (L_t)$ the side 
length in spatial (temporal) direction. The minimal distance
on the lattice, the so-called lattice spacing, is $a$ ($a_t$)
in space (time). This entails a maximum momentum on the lattice,
$p_{\rm max} = \pi/a$, which serves as an UV regulator of the theory.
The nucleons are point-like particles residing on the lattice sites,
whereas the nuclear interactions (pion exchanges and contact terms as described before
adapted to the lattice notation)
are represented as insertions on the nucleon world lines using standard
auxiliary field representations. The nuclear forces have an approximate
spin-isospin SU(4) symmetry (Wigner symmetry) \cite{Wigner:1936dx}
that is of fundamental importance in suppressing the malicious sign oscillations 
that plague any Monte Carlo (MC) simulation of strongly interacting fermion systems 
at finite density. 
\begin{figure}[t]
\begin{center}
\includegraphics[width=8cm]{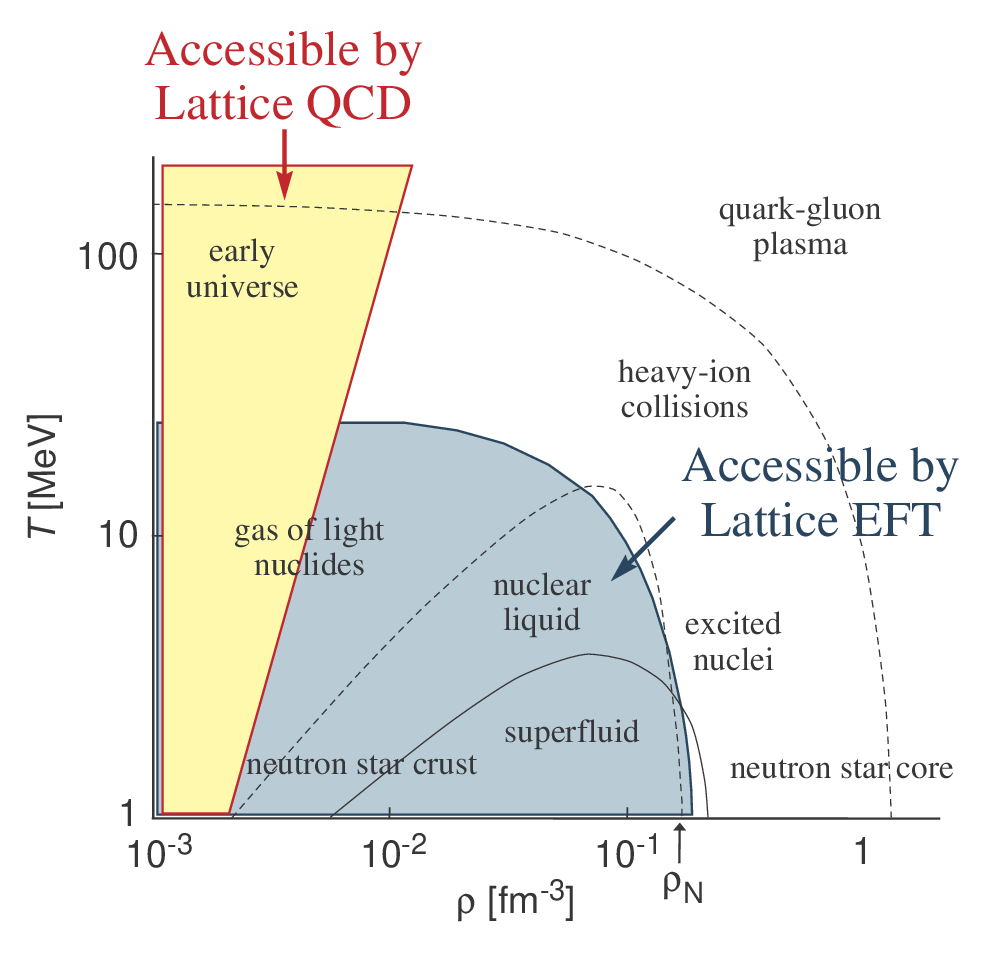}
\caption{\label{fig:phasedia}
Nuclear phase diagram as accessible by lattice QCD
and by nuclear lattice EFT. Figure courtesy of Dean Lee.
}
\end{center}
\end{figure}
For this reason, nuclear lattice simulations allow access to a large part
of the phase diagram of QCD, see Fig.~\ref{fig:phasedia}, whereas 
calculations using  lattice QCD are limited to finite temperatures and
small densities (baryon chemical potential). In what follows, I will
concentrate on the calculation of the ground state properties and excited
states of atomic nuclei with $A \leq 28$. 
The interactions of nucleons are simulated using the MC transfer projection
method. Each nucleon evolves as a single particle in a fluctuating
background of pion and auxiliary fields, the latter representing the
multi-nucleon contact interactions. One also performs Gaussian smearing of
the LO contact interactions which is required by the too strong binding
of four nucleons on one lattice site.  To leading order, one starts with a 
Slater determinant of single-nucleon standing waves in a periodic cube for 
$Z$ protons and $N$ neutrons (with $Z+N = A$) (or with more correlated
states, as described below). Further, the SU(4) 
symmetric approximation of the LO interaction is used as an approximate 
inexpensive filter for the first $t_0$
time steps -- this suppresses dramatically the sign oscillations. Then, one
switches on the full LO interaction and calculate the ground state energy and
other properties from the correlation function $Z(t) = \langle \Psi_A |\exp(-t
H)| \Psi_A \rangle$, letting the Euclidean time $t$ go to infinity. Higher order
contributions, the Coulomb repulsion between protons and other
isospin-breaking effects (due to the light quark mass difference) are computed as
perturbative corrections to the LO transfer matrix. This is symbolically 
depicted in Fig.~\ref{fig:time}.
Excited states are calculated from a multi-channel projection MC method.
As a first step, various improvements in our LO lattice action are used,
including ${\cal O}(a^4)$ improvements for the nucleon kinetic energy
and the Gaussian smearing factors of the contact interactions. Moreover,
all lattice operators at ${\cal O}(Q^3)$ are included, in particular
also the ones related to the breaking of rotational symmetry. Their
strengths can be tuned to eliminate unphysical partial wave mixing
like e.g. between the $^3S_1$-$^3D_1$ and the $^3D_3$ partial waves.
Most of the results presented below have been obtained with the following lattice set-up: 
$a = 1.97\,$fm, $N = 7$, $a_t = 1.32\,$fm.
The forces have been obtained at NNLO, with nine two-nucleon 
LECs fixed from fits to S- and P-wave $np$ phase shifts,
two isospin-breaking NN LECs determined from the $nn$ and $pp$ 
scattering lengths and two 3N LECs fixed  from 
the triton binding energy and the axial-vector contribution to triton 
$\beta$-decay~\cite{Gazit:2008ma}.
Further, there is some smearing required in the LO S-wave four-nucleon terms with 
its size parameter determined from the
average $np$ S-wave effective range~\cite{Borasoy:2006qn}.

\begin{figure}[!htb]
  \centering
  \includegraphics[width=0.45\textwidth]{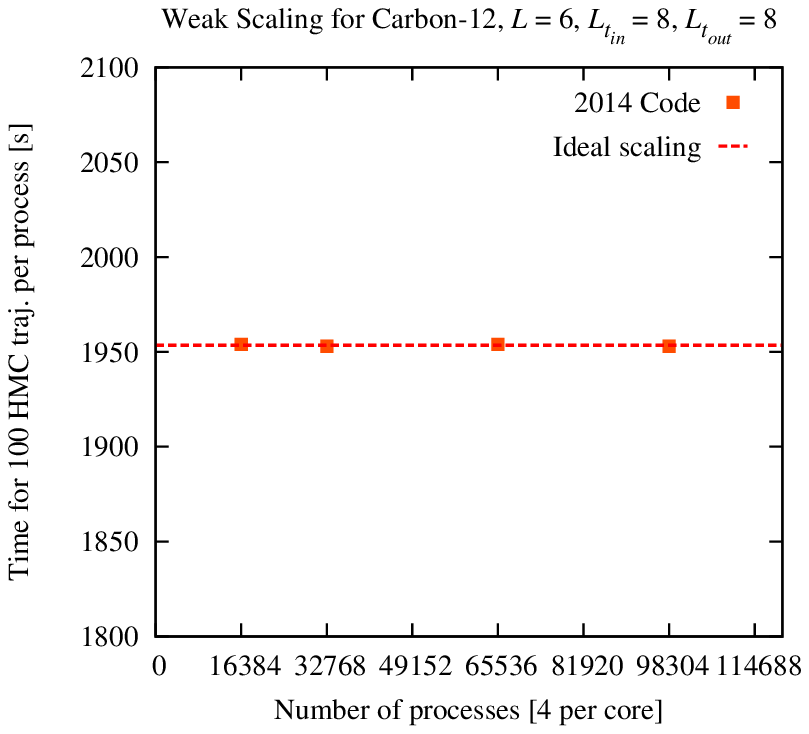}
    \caption{\footnotesize %
            Weak scaling of the NLEFT MC code with the number of processors. Here,
            weak scaling is defined as how the solution time varies with the number 
            of processors for a fixed problem size per processor.}
  \label{B9:time_procs}
\end{figure}

The world-line approach that maps the $A$ nucleon problem on the evolution
of $A$ independent particles (except for antisymmetrization) is perfectly
suitable for high-performance computer applications. In fact,
the low memory and extremely parallel structure of the lattice Monte
Carlo codes allow jobs on large parallel machines to
run very efficiently with hundreds of thousand processes.  We are able to run 
very efficiently with four processes per core on  the supercomputer JUQUEEN 
at Forschungszentrum J\"ulich, with very little loss 
in performance when compared  with one process per core, thereby achieving 
a factor of four increase in the total performance. 
In Fig.~\ref{B9:time_procs}  the
computational time for each process on the JUQUEEN supercomputer to produce 100
Hybrid Monte Carlo trajectories is shown. The time is plotted as a function of the
number of parallel processes with four processes per core. One sees that the performance is entirely
independent of the number of processes. 
The computational time for each JUQUEEN process to generate 100
Hybrid Monte Carlo trajectories versus the number of nucleons $A$  
scales as $79.7A + 7.11A^2$ for these values of $A$.  For smaller values the scaling is 
close to linear in $A$, while  the quadratic dependence becomes more important for larger $A$. The data
shown is for lattice simulations of $^4$He, $^8$Be, $^{12}$C, $^{16}$O, and $^{20}$Ne 
in a periodic cube with length $L=13.8$~fm and lattice spacing $a = 1.97\,$fm.
We are presently exploring a range of lattice spacings from $a \simeq 1\,$fm to
$a\simeq 2\,$fm to get a better handle on the discretization errors. 
\\


\section{Results form lattice simulations }
\label{sec:res}

Having discussed the framework of nuclear lattice simulations, we are now in the position
to present some results in this approach, but without going into any details. The interested
reader is referred to the original publications for more details. The works have been
done under the umbrella of the NLEFT collaboration\footnote{NLEFT stands for Nuclear
Lattice Effective Field Theory.} making use of supercomputing resources at 
Forschungszentrum J\"ulich and RWTH Aachen.

\smallskip

\begin{figure}[h!]
\begin{center}
\includegraphics[width=.27\textwidth]{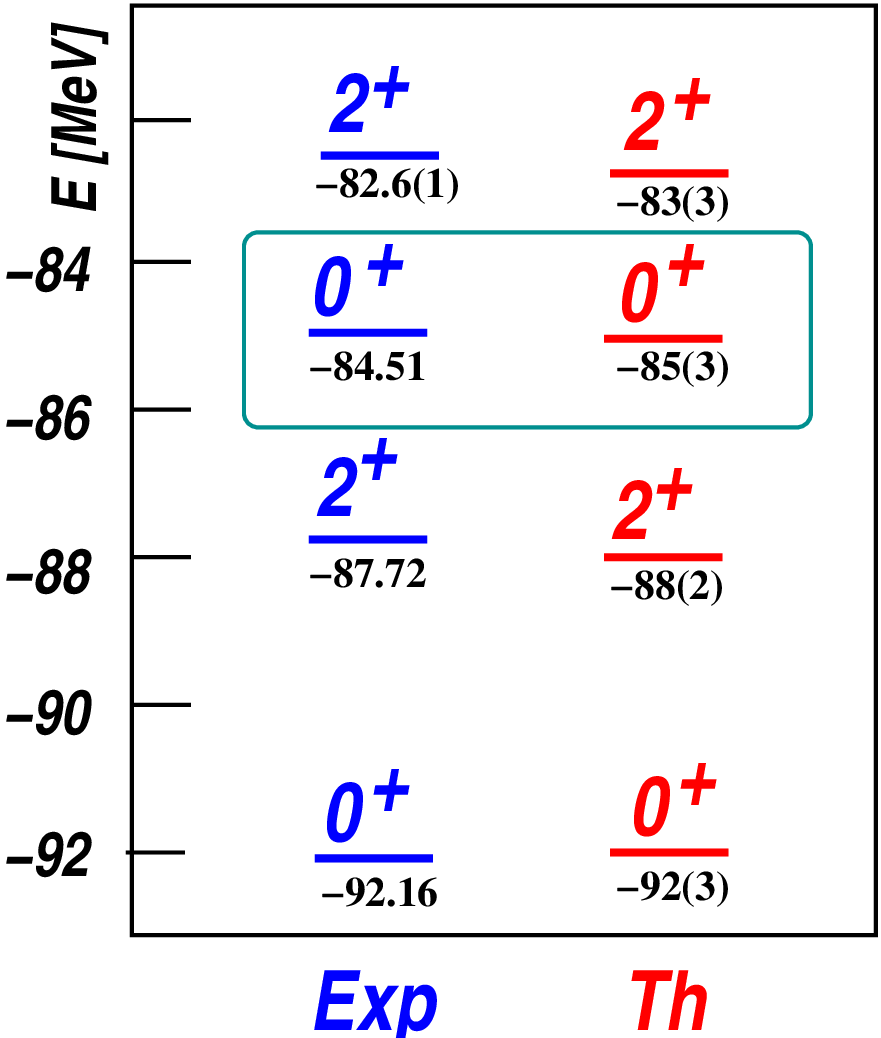}
\end{center}
\caption{Low-lying even-parity spectrum of $^{12}$C~\cite{Epelbaum:2011md,Epelbaum:2012qn}. 
The Hoyle state, the first excited spin zero, positive parity ($J^P = 0^+$) state 
is accentuated by the box. All
energies in MeV. ``Exp'' refers to the experimental values, while the NLEFT
calculation give the theoretical numbers denoted by ``Th''.
Note that these are {\em absolute} values, not just excitation
energies. Note further that by now the theoretical uncertainty given in the square
brackets has been decreased by about one order of magnitude.
\label{fig:spec12C}}
\end{figure}

\noindent{\em Ab initio calculation of the Hoyle state and its 
structure:}\\
The excited state of the $^{12}$C nucleus with $J^P = 0^+$
 known as the ``Hoyle state'' constitutes one 
of the most interesting, difficult and timely challenges in nuclear physics, as 
it plays a key role in the production of carbon via fusion of three alpha particles 
in red giant stars. The first {\em ab initio} calculation of the spectrum of
$^{12}$C was performed in Ref.~\cite{Epelbaum:2011md}, giving 
its first  $0^+$ excitation -- the Hoyle state -- at the proper energy, cf.
Fig.~\ref{fig:spec12C}.
This can be considered the breakthrough investigation for the method of
nuclear lattice simulations. In Ref.~\cite{Epelbaum:2012qn}, ab initio 
lattice calculations
were presented which unravel the structure of the Hoyle state, along with evidence 
for a low-lying spin-2 rotational excitation. For the $^{12}$C ground state and the 
first excited spin-2 state, we find a compact triangular configuration of alpha 
clusters. For the Hoyle state and the second excited spin-2 state, we find a 
``bent-arm'' or obtuse triangular configuration of alpha clusters. The calculated 
electromagnetic transition rates between the low-lying states of $^{12}$C have also 
been obtained at LO (higher order corrections still require improved codes). 

\smallskip

\begin{figure}[t]
\begin{center}
\includegraphics[width=8cm]{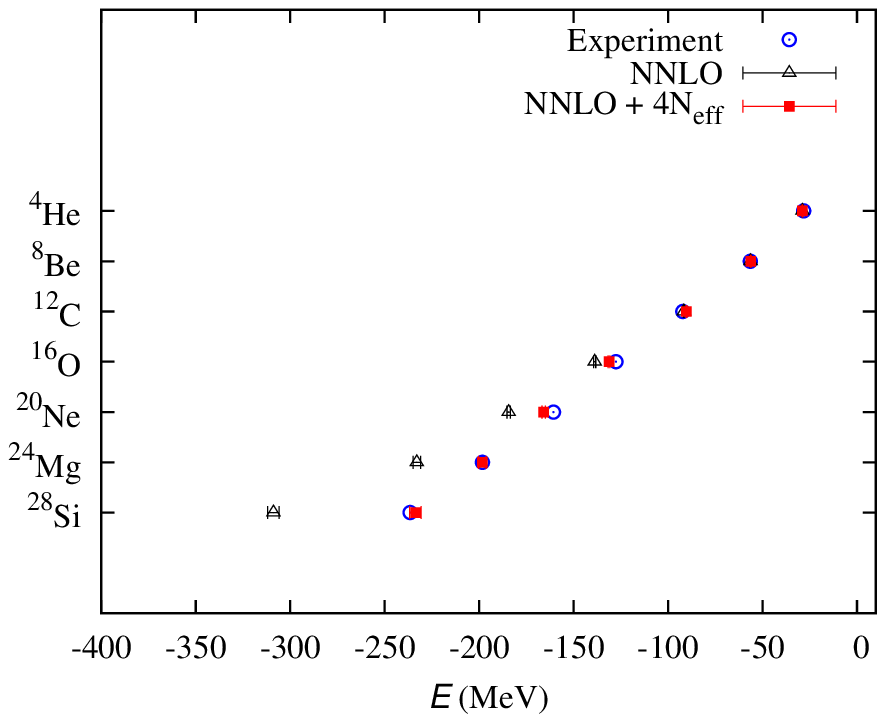}
\caption{\label{fig:alpha}
Ground-state energies of the alpha-cluster nuclei from $A=4$ to $A=28$.
The NNLO calculation is represented by the black triangles. The red squares
show the results including an effective 4N interaction, and the blue circles
are the experimental values. Figure courtesy of Dean Lee.
}
\vspace{-8mm}

\end{center}
\end{figure}
\noindent{\em Towards medium-mass nuclei}:\\
We have also extended nuclear lattice simulations 
to the regime  of medium-mass nuclei~\cite{Lahde:2013uqa}. To achieve that, a method which allows 
to greatly  decrease the uncertainties due to extrapolation at large Euclidean 
time was implemented. It is based on triangulation of the large Euclidean time
limit from a variety of SU(4) invariant initial interactions. 
The ground states of alpha nuclei from $^4$He to $^{28}$Si are 
calculated up to next-to-next-to-leading order in the EFT expansion.
With increasing atomic number $A$, one finds a growing overbinding as
shown in Fig.~\ref{fig:alpha}. Such effects are genuine to soft NN interactions
and also observed in other many-body calculations, 
see e.g. Refs.~\cite{Hagen:2012fb,Jurgenson:2013yya,Roth:2011ar}.
While the long-term objectives of NLEFT  are a decrease in the 
lattice spacing and the inclusion of higher-order 
contributions, it can be shown that the missing physics at NNLO can be approximated 
by an effective four-nucleon interaction. Fitting its strength to the binding
energy of $^{24}$Mg, one obtains an overall excellent description as depicted 
in Fig.~\ref{fig:alpha}. 

\smallskip

\noindent{\em Spectrum and structure of $^{16}$O}:\\
We have also performed  lattice calculations of the low-energy 
even-parity states of $^{16}$O~\cite{Epelbaum:2013paa}, which is another
mysterious alpha-cluster type nucleus. We find good agreement with the empirical 
energy spectrum, cf. Tab.~\ref{tab_en}, and with the electromagnetic 
properties and transition rates (after rescaling with the corrected charge
radius as detailed in \cite{Epelbaum:2013paa}). 
For the ground state, we find that the nucleons are arranged in a 
tetrahedral configuration of alpha clusters. For the first excited spin-0 state, 
we find that the predominant structure is a square configuration of 
alpha clusters, with rotational excitations that include the first spin-2 state. 
\begin{table}[h]
\begin{center}
\begin{tabular}{| c | r | r r | r |}
\hline 
$J_n^p$ & \multicolumn{1}{c |}{LO } & \multicolumn{1}{c}{NNLO} 
& \multicolumn{1}{c |}{+4N$_\mathrm{eff}$} &
\multicolumn{1}{c |}{Exp} \\ \hline\hline
$0^+_1$ & $-147.3(5)$  & $-138.8(5)$ & $-131.3(5)$ & $-127.62$ \\
$0^+_2$ & $-145(2)$ & $-136(2)$ & $-123(2)$ & $-121.57$ \\
$2^+_1$ & $-145(2)$ & $-136(2)$ & $-123(2)$ & $-120.70$ \\
\hline
\end{tabular}
\end{center}
\caption{NLEFT results and experimental (Exp) values for the lowest
  even-parity states of $^{16}$O (in MeV). 
The errors are one-standard-deviation estimates which include both statistical Monte Carlo errors and 
uncertainties due to the extrapolation $N_t^{} \to \infty$.
The combined statistical and extrapolation errors are given in
parentheses. The 
columns labeled ``LO'' and ``NNLO'' show
the energies at each order. Finally, the column ``+4N$_\mathrm{eff}$'' includes the effective 
4N contribution as discussed before.  
\label{tab_en}}
\end{table}

\pagebreak 

\noindent{\em Alpha clustering in  nuclei:}\\
$\alpha$-clustering is known to play an important role in the carbon nucleus $^{12}$C, 
see e.g. Refs.~\cite{Tohsaki:2001an,Bijker:2002ac,Chernykh:2007zz,Freer:2014qoa}, 
as well as in the oxygen nucleus  $^{16}$O, see e.g. 
Refs.~\cite{Robson:1979zz,Bauhoff:1984zza,Tohsaki:2001an,Bijker:2014tka,Freer:2005ia}. For other
recent work on alpha clustering also in heavier nuclei, 
see e.g. Refs.~\cite{Zhao:2014vfa,Ebran:2014pda}.
In nuclear lattice simulations of these nuclei, clustering emerges naturally. This is
due to the fact that the path integral samples all possible configurations, in particular
also the ones with four nucleons on one lattice site, as allowed by Fermi statistics.
The alpha cluster configurations in  $^{12}$C and in $^{16}$O can be obtained in two ways. 
First, one can prepare cluster-type initial states, say three alphas for $^{12}$C or
four alphas for  $^{16}$O,  and then investigate the time evolution of such cluster 
configurations and  extract e.g. the corresponding energies as the Euclidean time 
goes to infinity.  Second, one can also   start with initial states that have no clustering 
at all, the Slater determinants of standing waves mentioned before. One can then measure 
the four-nucleon correlations. For such initial states, this density grows quickly
 with time and reaches a high level. For the cluster initial states, these correlations 
start out at a high level and stay large as a function of Euclidean 
 time. This is a clear indication that the observed clustering is not built 
in by hand but rather follows from the strong four-nucleon correlations in 
 the considered nuclei. Or, stated differently, if one starts with an initial wave function
without  any clustering, on a short time scale clusters will form and make up the
 most important contributions to the structure of  $^{12}$C and $^{16}$O or any 
such type of nucleus, where $\alpha$-clustering is relevant. So the mysterious phenomenon
of $\alpha$-clustering emerges naturally in this novel approach to exactly solve the
nuclear $A$-body problem.

\smallskip

\noindent{\em A method to go beyond alpha-cluster nuclei:}\\So far,
we have performed  Projection Monte Carlo calculations of nuclear lattice EFT 
that suffer from sign  oscillations to a varying degree dependent on the number of protons and
neutrons.  Hence, such studies have hitherto been concentrated on nuclei with 
equal numbers of protons and neutrons, and especially on the alpha nuclei
where the sign oscillations are smallest. In Ref.~\cite{Lahde:2015ona},
we have introduced the technique of  ``symmetry-sign extrapolation'' which allows 
us to use the approximate Wigner SU(4) symmetry of the nuclear interaction to 
control the sign oscillations without introducing unknown systematic errors. 
The method can briefly be described as follows:  One defines the ``interpolating Hamiltonian'' $H$ as
\begin{equation}
H =  d_h^{} H_\mathrm{LO}^{} + (1-d_h^{}) H_{\rm SU(4)},
\label{Hd}
\end{equation}
which depends on the real parameter $d_h^{}$ as well as the (unphysical) coupling constant $C_{\rm SU(4)}^{}$ of
the SU(4) symmetric Hamiltonian $H_{\rm SU(4)}$.
This can also be viewed as giving the interaction parameters a linear dependence on $d_h^{}$.
By taking $d_h^{} < 1$, we can always decrease the sign problem to a tolerable level, while simultaneously tuning 
$C_{\rm SU(4)}^{}$ to a value favorable for an extrapolation $d_h^{} \to 1$. Most 
significantly, we can make use of the constraint that the physical result at $d_h^{} = 1$ should be independent of $C_{\rm SU(4)}^{}$. 
The dependence of calculated matrix elements on $d_h^{}$ is smooth in the vicinity of $d_h^{} = 1$. 
We have benchmarked this method by calculating the ground-state energies of 
the $^{12}$C, $^6$He and  $^6$Be nuclei. In the future, it will allow for 
studies of neutron-rich halo nuclei and asymmetric nuclear matter as well as exploring the limits of nuclear stability.
For a different extrapolation method used  in Shell Model Monte Carlo calculations about two decades ago, 
see Ref.~\cite{Koonin:1996xj}.

\smallskip

\begin{figure}[h!]
\begin{center}
\includegraphics[width=.45\textwidth]{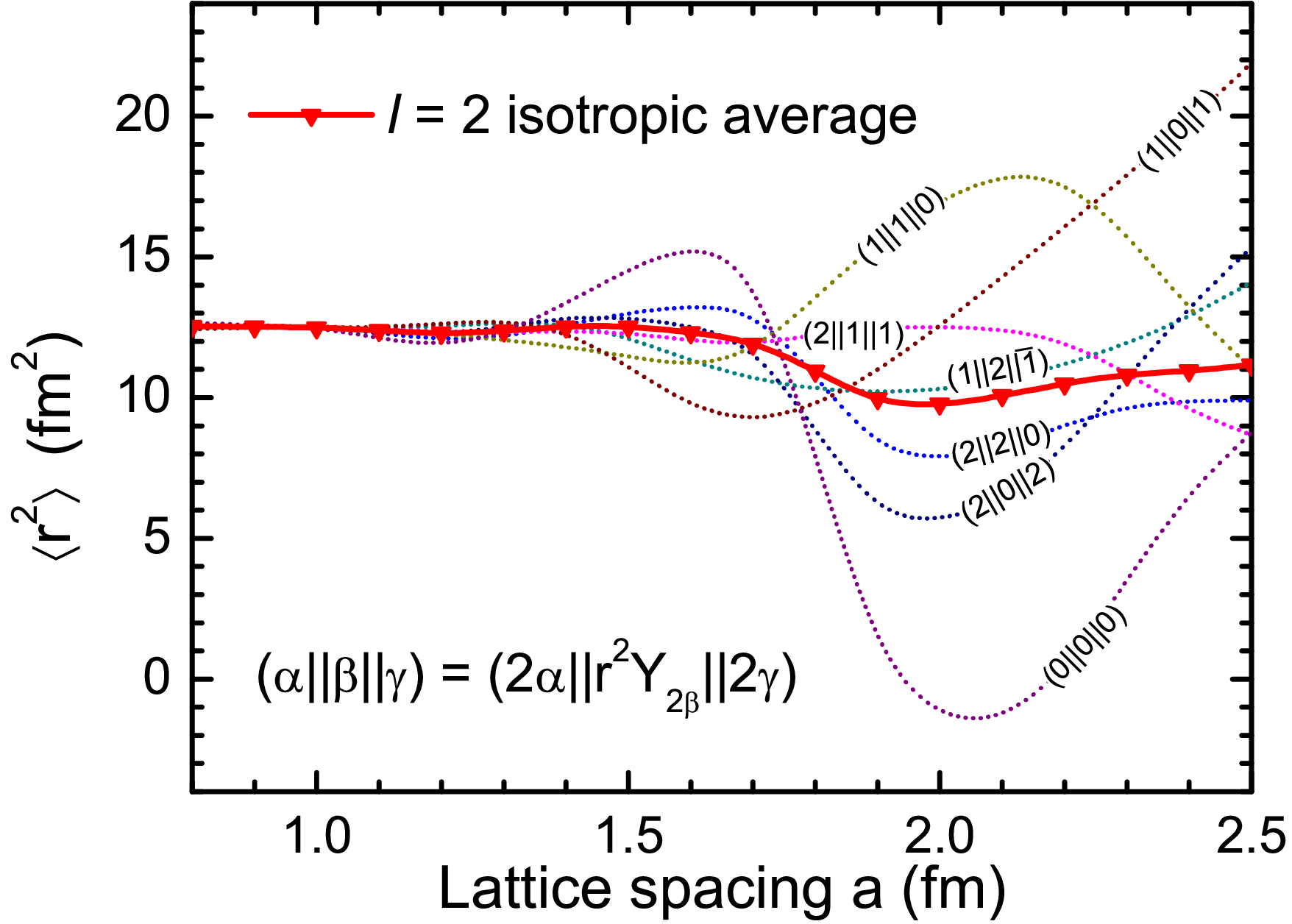}
\end{center}
\caption{Mean square radii $\langle r^2 \rangle$ for the lowest $2^+$ multiplet of $^{8}$Be states within a simplified alpha-cluster model calculation. The reduced lattice matrix elements all
merge in the limit $a \to 0$, while at finite $a$ the matrix elements depend on the quantum number $\alpha$, $\beta$ and $\gamma$, which is indicative of rotational symmetry breaking.
Such effects are nearly eliminated in the isotropic average, especially when $a \leq 1.7$~fm.
\label{fig:rot_avg}}
\end{figure}

\noindent{\em Overcoming rotational symmetry breaking:}\\
On the lattice, rotational invariance is broken from the full SO(3) rotational 
group to the cubic group. Hence, observables computed on the lattice will in general 
be affected by rotational symmetry breaking effects. In particular, the unambiguous 
identification of excited states and the computation of transition amplitudes may 
suffer significantly due to the  relatively large lattice spacings of $a \simeq 2$~fm 
in present nuclear lattice simulations. Hence, it makes sense to carefully determine 
the sources of rotational symmetry breaking in actual
NLEFT simulations, and search for methods that minimize their impact on physical 
observables. We have therefore used a simplified alpha cluster model to study the 
lattice matrix elements of irreducible tensor operators as a function of 
the lattice spacing $a$~\cite{Lu:2014xfa,Lu:2015gfa}.
In order to minimize the effects of rotational symmetry breaking, we have 
introduced the ``isotropic average'' which consists of a linear combination of the 
components of a given matrix element, such that each component is weighted according 
to the Clebsch-Gordan coefficient with the associated quantum numbers. This method, 
which is equivalent to averaging over all lattice orientations,
enables the unambiguous computation of matrix elements even at large lattice spacings. 
In Fig.~\ref{fig:rot_avg}, we illustrate the effect of isotropic averaging on the 
mean square radius of $^{8}$Be within the alpha cluster model. For related work in lattice
QCD, see e.g. Ref.~\cite{Davoudi:2012ya}.

\smallskip

\noindent{\em Lattice spacing dependence:}\\
As stated, most of the calculations of the NLEFT collaboration have been done
at the coarse lattice spacing $a \simeq 2\,$fm. Besides the studies of the lattice
spacing dependence in alpha cluster models just discussed, in Ref.~\cite{Klein:2015vna} 
we have investigated NLEFT for the two-body system 
for several lattice spacings ($0.5~{\rm fm} \leq a \leq 2 \,$fm) at lowest order 
in the pionless as well as in the  pionful theory. 
We find that in the pionless case, a simple Gaussian smearing allows 
to demonstrate lattice spacing independence over a wide range of lattice spacings. 
We show that regularization methods known from the continuum 
formulation~\cite{Epelbaum:2014efa} (as discussed in Sec.~\ref{sec:forces})
are necessary as well as feasible for the pionful approach. This leads to 
$a$-independent observables in the two-nucleon sector for the range of lattice
spacings mentioned.

\smallskip

The NLEFT collaboration is presently working out  next generation lattice forces that
have much reduced lattice artifacts and show a much reduced sign problem. This
will allow to substantially improve the precision of NLEFT calculations and will
give access to much larger $A$. Interesting times are ahead of us.


\section{Fine-tuning in nuclear physics and the anthropic principle}
\label{sec:anthro}

The elements that are pertinent to life on Earth are generated in the 
Big Bang and in stars through the fusion of protons, neutrons and nuclei.
In Big Bang nucleosynthesis (BBN), alpha particles and some
heavier elements are generated. Life essential elements like $^{12}$C and
$^{16}$O are generated in hot, old stars, where the so-called triple-alpha
reaction plays an important role. Here, two alphas fuse to produce the
instable, but long-lived $^8$Be nucleus. As the density of $^4$He nuclei in
such stars is high, a third alpha fuses with this nucleus before it decays.
However, to generate a sufficient amount of $^{12}$C and $^{16}$O, an excited
state in $^{12}$C at an excitation energy of 7.65~MeV with spin zero and
positive parity is required as pointed out by Hoyle long ago \cite{Hoyle} (which 
was already discussed briefly before).
In a further step, carbon is turned into oxygen without such a resonant
condition. So we are faced with a multitude of fine-tunings which need to
be explained. We  already know that  all strongly interacting composites like 
hadrons and nuclei must emerge from the underlying gauge theory of the
strong interactions, Quantum
Chromodynamics (QCD), that is formulated in terms of quarks and gluons.
These fundamental matter and force fields are, however, confined. Further,
the mass of the light quarks relevant for nuclear physics is very small
and thus plays little role in the total mass of nucleons and nuclei. Finally,
protons and neutrons form nuclei. This requires the inclusion of
electromagnetism, characterized by the fine-structure constant $\alpha_{\rm
  EM}\simeq 1/137$. So the question we want to address in the following is:
How sensitive are these strongly interacting composites to variations in the
fundamental parameters of QCD+QED? The role of the weak interactions
is more subtle, see the later discussion and also the interesting paper~\cite{Harnik:2006vj}.

\begin{figure}[t]
\begin{center}
\includegraphics[width=.475\textwidth]{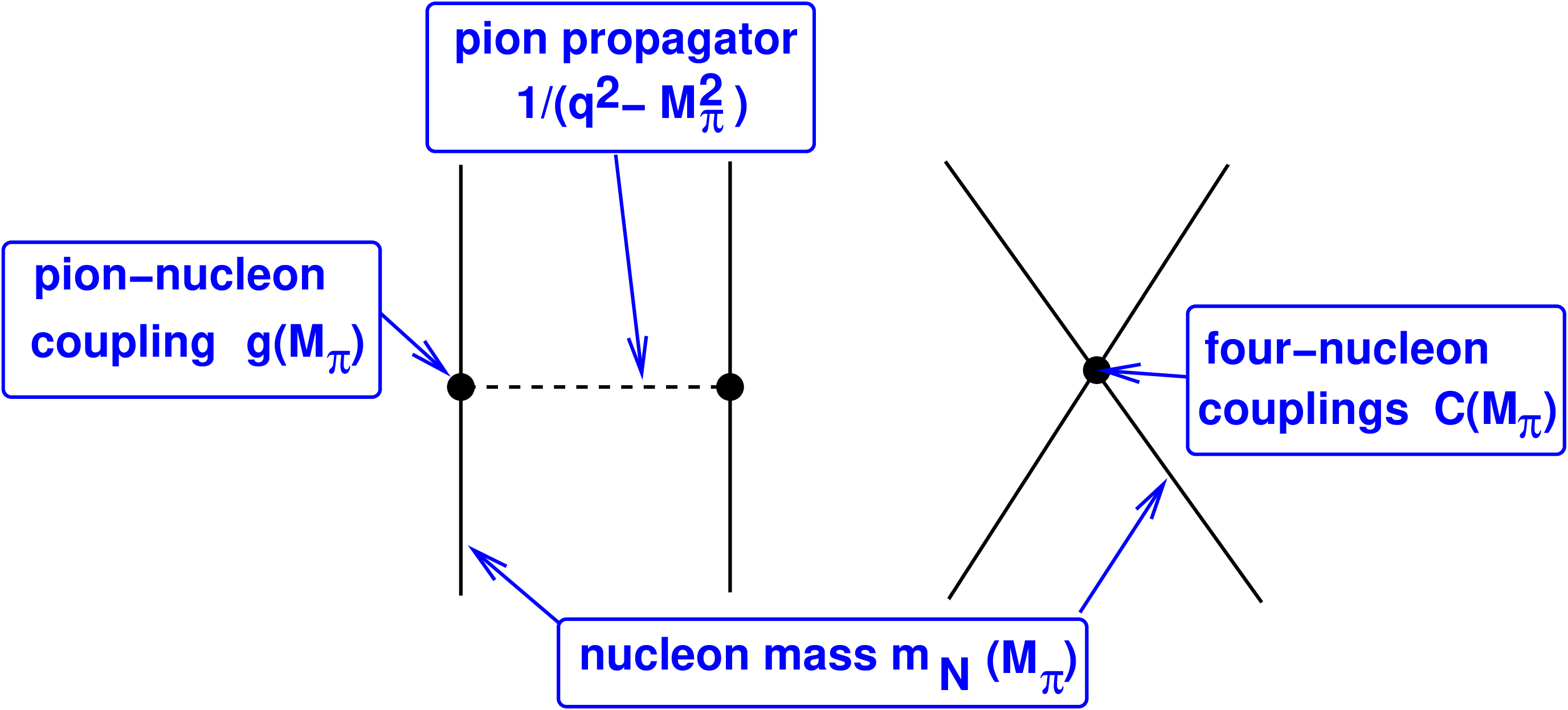}
\end{center}
\caption{Explicit and implicit pion (quark) mass dependence of the
leading order nucleon-nucleon (NN) potential. Solid (dashed) lines denote
nucleons (pions).}
\label{fig:mpi}
\end{figure}

First, let me discuss the fine-tunings related to the strong interactions.
In the Weinberg scheme discussed so far, the quark mass
dependence of the forces is generated explicitly (through the pion propagator) and
implicitly (through the pion-nucleon coupling, the nucleon mass, and the four-nucleon 
couplings), see  Fig.\ref{fig:mpi}.

Throughout, we use the Gell-Mann--Oakes--Renner relation~\cite{GellMann:1968rz}, 
$M_\pi^2 = B(m_u+m_d)$, so one can use the notions pion and quark mass dependence synonymously.
 For any observable ${\cal O}$ of a hadron $H$, we can define
its quark mass dependence in terms of the so-called $K$-factor, 
$\delta {\cal O}_H/\delta m_f \equiv K_H^f \,({\cal O}_H/m_f)$, with $f=u,d,s$,
and $m_f$ the corresponding light quark mass.
The pion mass dependence of pion and nucleon properties can be obtained from
lattice QCD combined with chiral perturbation theory as detailed in
Ref.~\cite{Berengut:2013nh}$\,$. The pertinent results are: 
$K_{M_\pi}^q = 0.494^{+0.009}_{-0.013}$, $K_{F_\pi}^q = 0.048\pm 0.012$, and
$K_{m_N}^q = 0.048^{+0.002}_{-0.006}$, where $q$ denotes the average light
quark mass. For the quark mass dependence of the short-distance terms, 
one has to resort to modeling using resonance saturation~\cite{Epelbaum:2001fm}. This induces
a sizeable uncertainty that might be overcome by lattice simulations in the
future. For the NN scattering lengths, this leads to $K^q_{1S0} = 2.3^{+1.9}_{-1.8}$,
$K^q_{3S1} = 0.32^{+0.17}_{-0.18}$ and $K^q_{\rm BE(deut)} =
-0.86^{+0.45}_{-0.50}$ (with BE denoting the binding energy), extending and 
improving earlier work based on EFTs and 
models, see e.g. Refs.~\cite{Muther:1987sr,Beane:2002vs,Epelbaum:2002gb,Flambaum:2007mj,Soto:2011tb}.
We point out the recent work of Ref.~\cite{Baru:2015ira}, which derives low-energy theorems
for nucleon-nucleon scattering at unphysical quark masses and relates to the recent 
lattice QCD calculations at large pion masses~\cite{Beane:2013br}. In view of these
new results, a re-evaluation of the $K$-factors should be done.
In the chiral EFT considered here, effects of shifts in $\alpha_{\rm EM}^{}$, that is modifications
of the electromagnetic interactions, can also be calculated.

With these results, we are now in the position to analyze what
constraints on possible quark mass variations the element abundances in BBN imply.
To answer this question, we also need the variation of $^3$He and $^4$He
with the pion mass. Following Ref.~\cite{Bedaque:2010hr} (BLP), these can be
obtained by convoluting the 2N $K$-factors with the variation of the
3- and 4-particle BEs with respect to the singlet and triplet NN scattering 
lengths. This gives $K_{^3{\rm He}}^q = -0.94\pm 0.75$ and  $K_{^3{\rm He}}^q 
= -0.55\pm 0.42$~\cite{Berengut:2013nh}, which is consistent with a direct
calculation using nuclear lattice simulations, $K_{^3{\rm He}}^q = -0.19\pm
0.25$ and $K_{^3{\rm He}}^q = -0.16\pm 0.26$~\cite{Lahde}. With this input,
we can calculate the BBN response matrix of the primordial abundances $Y_a$ at 
fixed baryon-to-photon ratio.  Comparing  the calculated with the observed
abundances, one finds that the most stringent limits arise from the
deuteron abundance [deut/H] and the $^4$He abundance normalized to the one
of protons, $^4$He($Y_p$), as most neutrons end up in alpha particles.
This  leads to the constraint {$\delta m_q/m_q = (2\pm 4)\%$}.
In contrast to most earlier determinations, we provide reliable error
estimates due to the underlying EFT. However, as pointed out by BLP, one
can obtain an even stronger bound due to the neutron lifetime, which 
strongly affects  $^4$He($Y_p$). We have re-evaluated this constraint under
the model-independent assumption that {\em all} quark and lepton masses
vary with the Higgs vacuum expectation value $v$, leading to 
\begin{equation}
\left|\frac{\delta v}{ v} \right| = \left| \frac{\delta m_q}{m_q}\right| \leq 0.9\%~. 
\end{equation}

\begin{figure}[t]
\begin{center}
\includegraphics[width=.45\textwidth]{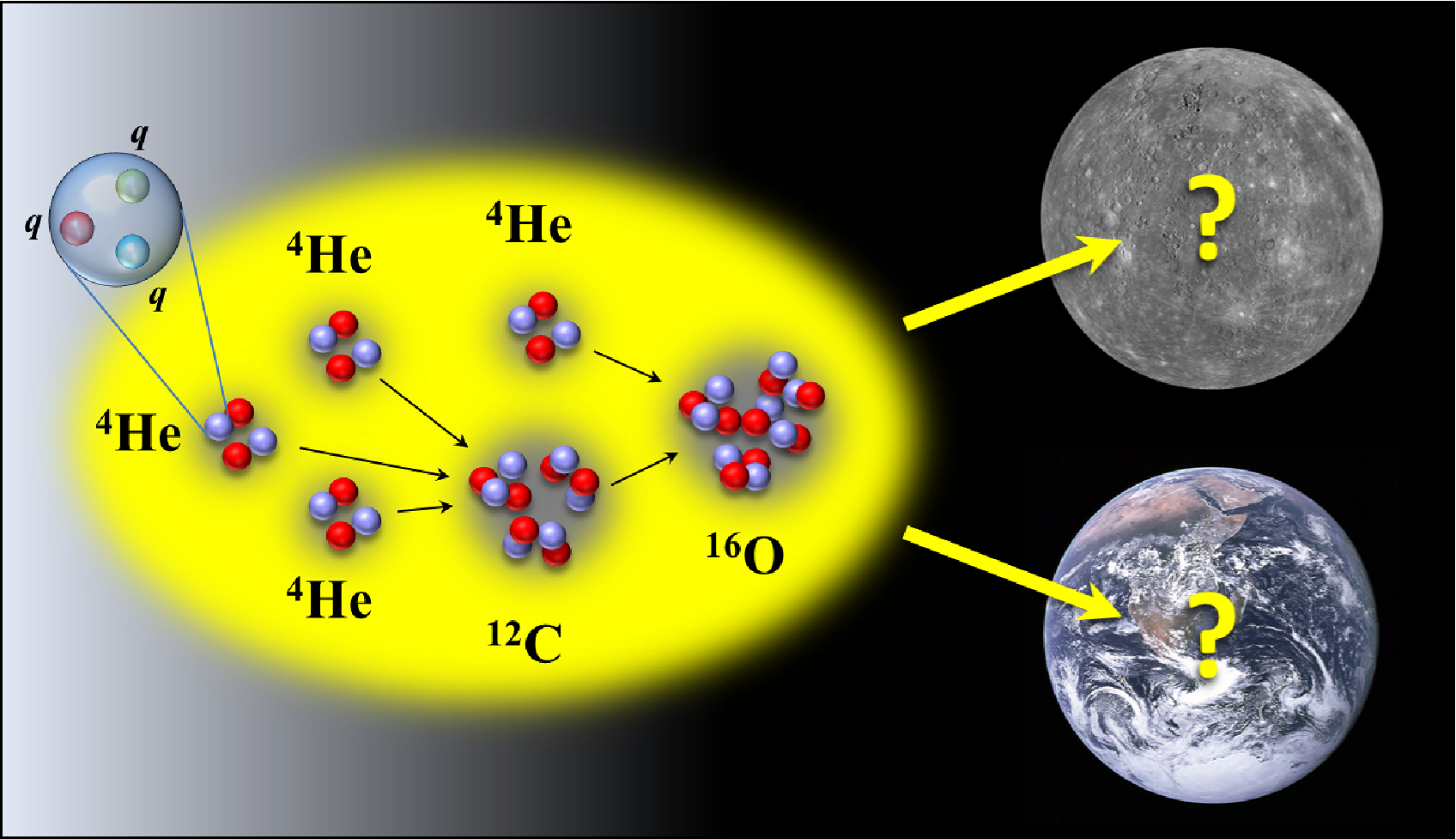}
\end{center}
\caption{Graphical representation of the question of how fine-tuned is
life on Earth under variations of the average light quark mass and $\alpha_{\rm EM}$. 
Figure courtesy of Dean Lee. }
\label{fig:fate}
\end{figure}

\begin{figure}[t]
\begin{center}
\includegraphics[width=.45\textwidth]{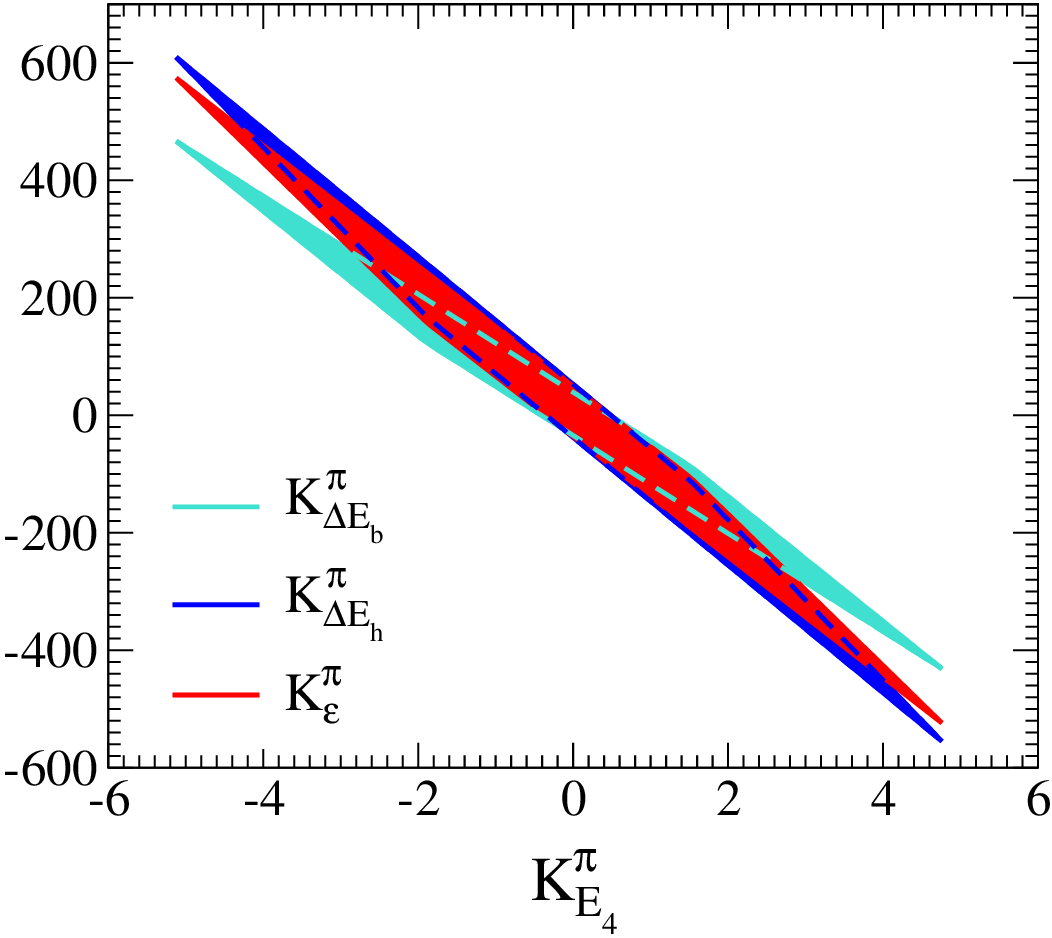}
\end{center}
\caption{Sensitivities of $\Delta E_h^{}$, $\Delta E_b^{}$ and $\varepsilon$ 
to changes in $M_\pi^{}$, as a function of $K_{E_4^{}}^\pi$ under independent 
variation of $\bar A_s^{}$ and $\bar A_t^{}$ over the range $\{-1 \ldots 1\}$.   
The bands correspond to $\Delta E_b^{}$, 
$\varepsilon$ and  $\Delta E_h^{}$ in clockwise order.
\label{fig:corr}}
\end{figure}

Next, let us consider the fine-tunings in the production of carbon and oxygen.
Stated differently, how much can we change  these
parameters from their physical values to still have an habitable Earth as shown
in Fig.~\ref{fig:fate}? To be more precise, we must specify which parameters
we can vary. In QCD, the strong coupling constant is tied to the nucleon mass
through dimensional transmutation. However, the light quark mass (here, only
the strong isospin limit is relevant) is an external parameter. Naively, one
could argue that due to the small contribution of the quark masses to the
proton and the neutron mass, one could allow for sizeable variations. However,
the relevant scale to be compared to here is the average binding energy per
nucleon, $E/A \leq 8\,$MeV (which is much smaller than the nucleon mass). 
As noted before, the Coulomb repulsion between protons is an important
factor in nuclear binding, therefore we must also consider changes in
$\alpha_{\rm EM}$.  Let us consider first fine-tunings in  QCD (for details, see
Refs.~\cite{Epelbaum:2012iu,Epelbaum:2013wla}). We want to calculate
the variations of the pertinent energy differences  in the
triple-alpha process $\delta \Delta E/ \delta M_\pi$, which according to
Fig.~\ref{fig:mpi} boils down to (we consider small variations around
the physical value of the pion mass $M_\pi^\mathrm{ph}$):
\begin{eqnarray}
\left. \frac{\partial E_i^{}}{\partial M_\pi^{}} \right|_{M_\pi^\mathrm{ph}} &= &
\left. \frac{\partial E_i^{}}{\partial \tilde M_\pi^{}} \right|_{M_\pi^\mathrm{ph}}
+ x_1^{} \left. \frac{\partial E_i^{}}{\partial m_N^{}} \right|_{m_N^\mathrm{ph}}
+ x_2^{} \left. \frac{\partial E_i^{}}{\partial \tilde g_{\pi N}^{}} \right|_{\tilde g_{\pi N}^\mathrm{ph}} 
\nonumber \\
& +& x_3^{} \left. \frac{\partial E_i^{}}{\partial C_0^{}}
\right|_{C_0^\mathrm{ph}}
+ x_4^{} \left. \frac{\partial E_i^{}}{\partial C_I^{}} \right|_{C_I^\mathrm{ph}},
\label{Eeq2}
\end{eqnarray}
with the definitions
\begin{align}
&& x_1^{} \equiv \left. \frac{\partial m_N^{}}{\partial M_\pi^{}} 
\right|_{M_\pi^\mathrm{ph}}, ~~
x_2^{} \left. \equiv \frac{\partial g_{\pi N}^{}}{\partial M_\pi^{}}
\right|_{M_\pi^\mathrm{ph}}~\nonumber\\
&& x_3^{} \equiv \left. \frac{\partial C_0^{}}{\partial M_\pi^{}}
\right|_{M_\pi^\mathrm{ph}}, ~~
x_4^{} \equiv \left. \frac{\partial C_I^{}}{\partial M_\pi^{}} \right|_{M_\pi^\mathrm{ph}},
\label{xy}
\end{align}
with $\tilde{M_\pi}$ the pion mass appearing in the pion-exchange potentials.
The various derivatives in Eq.~(\ref{Eeq2}) can be obtained precisely using our
Auxiliary Field Quantum Monte Carlo techniques and the $x_i$ ($i = 1,2,3,4$) are related to
the pion and nucleon $K$-factors just discussd.
The scheme-dependent quantities $x_{3,4}$ can be traded for the
pion-mass dependence of the inverse singlet and triplet scattering lengths,
\beq
\bar A_{s}^{} \equiv {\partial a_{s}^{-1}}/{\partial M_\pi^{}}|_{M_\pi^{\rm
    ph}}~,~~ \bar A_{t}^{} \equiv {\partial a_{t}^{-1}}/{\partial M_\pi^{}}|_{M_\pi^{\rm ph}}~.
\eeq
We can then
express all energy differences appearing in the triple-alpha process 
($\Delta E_b^{} \equiv E_8^{} - 2 E_4^{}, \Delta E_h^{} \equiv E_{12}^\star - E_8^{} - E_4^{},
\varepsilon =  E_{12}^\star - 3E_4^{}$, with $E_4^{}$ and $E_8^{}$ for the energies of 
the ground states of $^{4}$He and $^{8}$Be, respectively, and $E_{12}^\star$ denotes the
energy of the Hoyle state) as functions
of $\bar A_{s}$ and $\bar A_{t}$. One finds that all these energy differences
are correlated, i.e. the various fine-tunings in the triple-alpha process are not independent
of each other, see  Fig.~\ref{fig:corr}. Further, one finds a 
strong dependence on the variations of the
$^4$He BE, which is related to the $\alpha$-cluster structure of
the $^8$Be, $^{12}$C ground and Hoyle states.  Such correlations related to the production 
of carbon have indeed been speculated upon earlier~\cite{Livio,WeinbergFacing}.

\begin{figure}[h]
\begin{center}
\includegraphics[width=.45\textwidth]{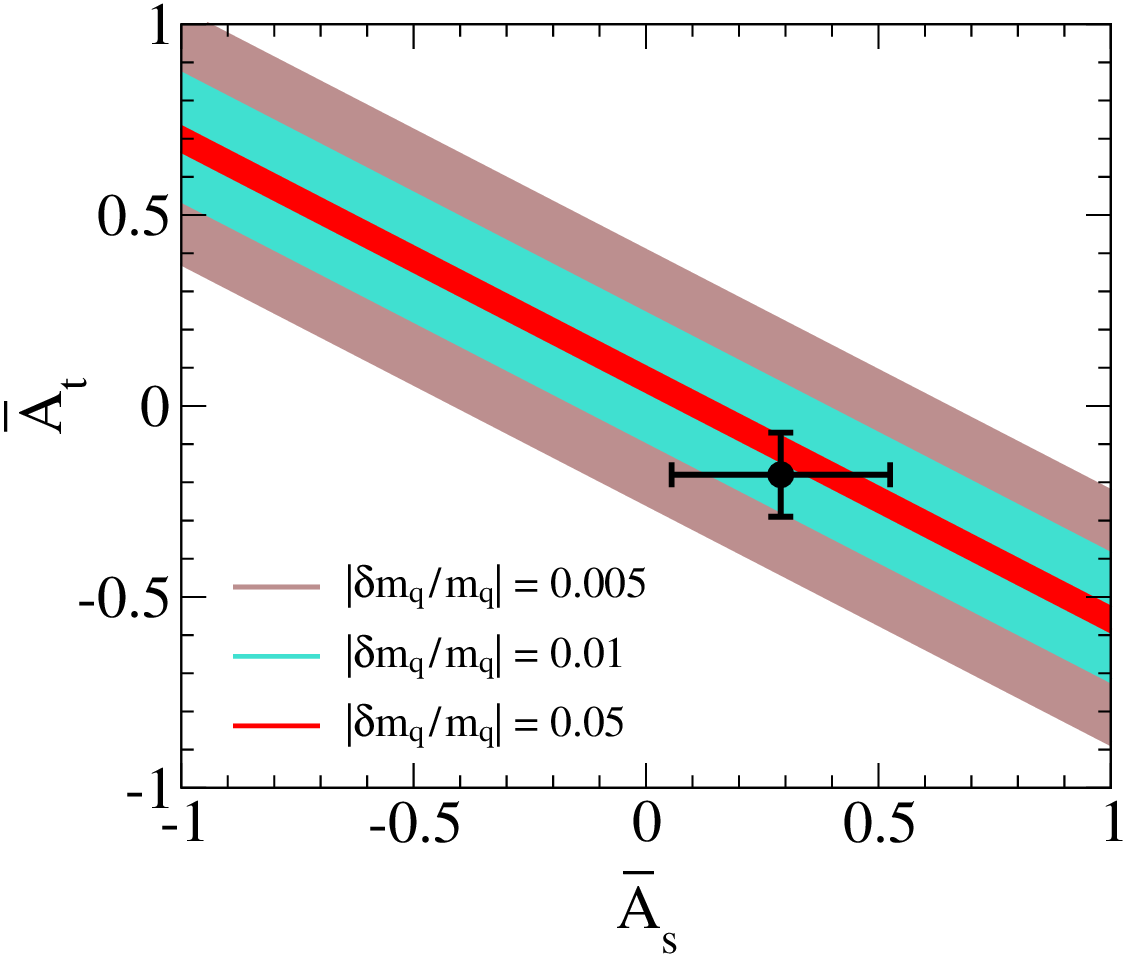}
\end{center}
\caption{``Survivability bands'' for carbon-oxygen based life from
  Eq.~(\ref{final_res}), due to  $0.5\%$ (broad outer band), $1\%$ (medium
  band) and $5\%$ (narrow inner band) changes in $m_q^{}$ in terms of the
  input parameters $\bar A_s^{}$ and $\bar A_t^{}$. The most up-to-date 
  N$^2$LO analysis of $\bar A_s^{}$ and $\bar A_t^{}$ from
  Ref.~\cite{Berengut:2013nh} is given by the data point with 
  horizontal and vertical error bars.}
\label{fig:band}
\end{figure}

Consider now the reaction rate of the triple-alpha process as given by
$r_{3 \alpha}^{} \sim N_\alpha^3 \Gamma_\gamma^{} \exp \left( -{\varepsilon}/{k_{\rm B}^{} T} \right)$,
with $N_\alpha$ the $\alpha$-particle number density in the stellar plasma with temperature $T$, 
$\Gamma_\gamma = 3.7(5)\,{\rm meV}$ the radiative width of the Hoyle state and 
$k_B$ is Boltzmann's constant.
The stellar modeling calculations of Refs.~\cite{Oberhummer-astro,Oberhummer:2000mn} suggest 
that sufficient abundances of both carbon and oxygen can be maintained within an envelope of 
$\pm 100$~keV around the empirical value of $\varepsilon = 379.47(18)$~keV. 
This condition can be turned into a constraint on shifts in $m_q^{}$ that reads
(for more details, see Ref.~\cite{Epelbaum:2013wla})
\beqa
&&\left| \Big[ 0.572(19) \, \bar A_s^{} + 0.933(15) \, \bar A_t^{} 
- 0.064(6)  \Big]  \left(\frac{\delta m_q^{}}{m_q^{}} \right) \right| 
 \nonumber\\
 && \qquad < 0.15\%~.
\label{final_res}
\eeqa
The resulting constraints on the values of $\bar
A_s^{}$ and $\bar A_t^{}$ compatible
with the condition $| \delta \varepsilon | < 100$~keV are visualized in 
Fig.~\ref{fig:band}.  The various shaded bands in this figure cover the 
values of $\bar A_s^{}$ and $\bar A_t^{}$ consistent
with carbon-oxygen based life, when $m_q^{}$ is varied by $0.5$\%, $1$\% and $5$\%.
Given the current theoretical 
uncertainty in $\bar A_s^{}$ and $\bar A_t^{}$, our results remain compatible 
with a vanishing $\partial \varepsilon / \partial M_\pi^{}$, in other words
with a complete lack of fine-tuning. Interestingly, Fig.~\ref{fig:band}  
also  indicates that the triple-alpha process is unlikely to be fine-tuned
to a higher degree than $\simeq 0.8$\% under variation of $m_q^{}$. 
The central values of $\bar A_s^{}$ and $\bar A_t^{}$ from
Ref.~\cite{Berengut:2013nh} suggest that variations in the light quark masses 
of up to $2 - 3$\% are unlikely to be catastrophic to the formation of 
life-essential carbon and oxygen. A similar calculation of the tolerance for 
shifts in the fine-structure constant $\alpha_{\rm EM}^{}$
suggests that carbon-oxygen based life can withstand shifts of 
$\simeq 2.5$\% in $\alpha_{\rm EM}^{}$. 

Finally, let me review the consequences of these results for our anthropic
view of the Universe.  As it is well known, the Hoyle state dramatically increases the reaction rate of the
triple-alpha process. The resulting enhancement
is also  sensitive to the exact value of $\varepsilon$, which is
therefore the principal control parameter of
this reaction. As the Hoyle state is crucial to the formation of elements
essential to life as we know it, this state
has been nicknamed the ``level of life''~\cite{Linde}. 
Thus, the Hoyle state is often viewed as a prime manifestation of the anthropic
principle, which states that the observable 
values of the fundamental physical and cosmological parameters are restricted
by the requirement that life can form to 
determine them, and that the Universe be old enough for that to
occur~\cite{Carter,Carr}. See, however, Ref.~\cite{Kragh} for a thorough historical
discussion of the Hoyle state in view of the anthropic principle\footnote{Many physicists consider
the anthropic principle with suspicion or disguise. However, one has to realize that one can derive
physics questions from it, like the ones discussed here. And it is certainly one important goal of
theoretical physics to explore alternative worlds in which fundamental parameters take different values
than the ones in Nature.}.
We remark that in the context of cosmology and string 
theory, the anthropic principle and its consequences have had a significant influence,
as reviewed recently in \cite{Schellekens:2013bpa}.
As noted already in Ref.~\cite{WeinbergFacing}, the allowed variations in $\varepsilon$
are not that small, as $|\delta \varepsilon/\varepsilon | \simeq 25\%$ still allows
for carbon-oxygen based life. So one might argue that the anthropic principle is
indeed {\it not} needed to explain the fine-tunings in the triple-alpha process.
However, as we just showed, this translates into allowed quark mass variations
of $2-3\%$ and modifications of the fine-structure constant of about 2.5\%.
The  fine-tuning in the fundamental parameters is thus much more severe than the
one in the energy difference $\varepsilon$. Therefore, beyond such relatively small changes 
in the fundamental parameters, the anthropic principle indeed appears necessary to 
explain the observed abundances of $^{12}$C and $^{16}$O.  A more detailed account of these
considerations is given in the review~\cite{Meissner:2014pma}.


\section{Nuclei as precision laboratories}
\label{sec:precision}

In this section, I show how chiral EFT can be used to test physcis within and
beyond the Standard Model (SM). I focus here on hadronic parity violation,
the calculation of light ion electric dipole moments to test CP violation,
and the use of chiral EFT currents to perform better calculation for WIMP
scattering off nuclei, that is used for direct dark matter detection.
The first two topics have recently been reviewed~\cite{deVries:2015gea},
so I only discuss the underlying physics and results obtained in the last few years.

The observation of parity violation (PV) in the weak interaction is one of
the pillars on which the SM of particle physics was built. 
In the SM, PV is induced because only left-handed quarks and leptons 
participate in the (charged current) weak interaction. At the fundamental level, 
parity violation originates from the exchange of the charged (and neutral)  weak gauge bosons. 
For low-energy (hadronic) processes, the heavy gauge bosons decouple from the theory leading to 
effective PV four-fermion interactions. The effective interactions 
resulting from the exchange of charged gauge bosons induce, for example, the 
beta-decay of the muon and the neutron, while the exchange of both charged and neutral gauge 
bosons gives rise to various PV four-quark operators.  
Although PV induced by the weak interaction is well understood at the level of elementary quarks, its manifestation at the hadronic and nuclear level is not well understood. This holds particularly true for the strangeness-conserving part of the weak interaction which induces PV in hadronic and nuclear systems. The SM predicts PV forces between nucleons, however, their forms and strengths are masked by the nonperturbative nature of QCD at low energies. In order to circumvent this problem, the  NN
interaction has been parametrized in the past through PV meson exchanges with adjustable strengths, the so-called DDH-framework \cite{Desplanques:1979hn}.
Given enough experimental input the unknown couplings can be determined and 
other processes can then be predicted. However, the extractions of the DDH coupling
constants from different experiments seem to be in disagreement \cite{Haxton:2013aca,Schindler:2013yua}.

In the last years, we have developed a  framework for hadronic and nuclear PV based on chiral EFT and applied this  to calculate PV 
hadronic and nuclear observables.  This approach has a number of big advantages over the more traditional DDH model. First of all, 
there is a clear link to the underlying theory, \textit{i.e.}, QCD supplemented with PV four-quark operators. Second, the EFT approach 
makes it possible to calculate the $P$-even and -odd NN potentials within the same framework. The resulting potentials can then be treated on 
the same footing. Third, the chiral Lagrangian can be improved by going to higher orders in the expansion. In fact, we have developed the PV potential 
up to N$^2$LO. Fourth, the chiral approach can be extended to other systems, such as reactions involving photons, which require the calculation of PV currents. 
These currents can be evaluated within the same framework as the potential, something which is not possible in the DDH model 
where the currents need to be modeled separately (in principle, using the method of unitary transformations, see e.g. Ref.~\cite{Gari:1976kj}, 
this could be achieved but that approach has never been applied to this program). 

Already some work in the past has been done on deriving a chiral EFT PV potential.  
At leading order in the power counting, the only term appearing in the chiral Lagrangian is the weak pion-nucleon vertex 
\begin{equation}\label{PoddLO}
\mathcal L_{PV } = \frac{h_\pi}{\sqrt{2}} \bar N (\vec \pi\times \vec \tau)^3 N\,\,\,,
\end{equation}
proportional to the LEC  $h_\pi$ \cite{Kaplan:1992vj}. Together with the usual pseudovector $P$-conserving (PC)
pion-nucleon interaction, the LO PV potential follows as
\begin{equation}
V_{1\pi}
= - \frac{g_{A}h_\pi}{ 2\sqrt{2} F_\pi} i(\vec \tau_1\times \vec \tau_2)^3 \frac{(\vec \sigma_1+\vec \sigma_2)\cdot \vec q }{M_\pi^2+q^2}\,\,\,,
\label{onepion}
\end{equation}
in terms of the nucleon spin $\vec \sigma_{1,2}$ and the momentum transfer flowing from nucleon $1$ to nucleon $2$: $\vec q = \vec p - \vec p^{\,\prime}$ ($q = |\vec q\,|$), where $\pm \vec p$ and $\pm \vec
p^{\,\prime}$ are the momenta of the incoming and
outgoing nucleons in the center-of-mass frame. The LO OPE potential changes the total isospin of the interacting nucleon pair and, at low energies, dominantly contributes to the ${}^3 S_1\leftrightarrow {}^3 P_1$ transition. The isospin change ensures that the LO potential vanishes for $pp$ and $nn$ scattering.
At NLO the number of LECs proliferates. First of all, five NN short-range contact interactions appear.  In addition, there appear two-pion-exchange (TPE) diagrams that are proportional to $h_\pi$ just as the OPE potential \cite{Zhu:2004vw,Kaiser:2007zzb}.
However, in contrast to the OPE potential, the TPE potential does contribute to PV in $pp$ scattering. This leads to a dependence on $h_\pi$ on PV $pp$ observables which had not been taken into account in earlier studies.  In fact, one of the main goals of experiments on hadronic and nuclear PV is to measure the size of $h_\pi$. At the moment the size of $h_\pi$ is unknown despite several measurements of  nuclear PV observables. Several theoretical estimates exist in the literature:
The most simple one is the use of naive-dimensional analysis  (NDA) which gives the following estimate
$h_\pi \sim \mathcal O(G_F F_\pi \Lambda_\chi) \sim 10^{-6}$, 
in terms of the Fermi coupling constant $G_F$. In the original DDH paper \cite{Desplanques:1979hn}, the authors have attempted to 
estimate $h_\pi$ using SU(6) symmetry arguments and the quark model finding the reasonable range $0\leq h_\pi \leq 1.2\cdot 10^{-6},$
and a ``best'' value of $h_\pi \simeq 4.6 \cdot 10^{-7}$, consistent with the 
NDA estimate.  The authors of Ref. \cite{Kaiser:1989fd} have calculated several PV 
meson-nucleon vertices in a framework of a non-linear chiral Lagrangian where the 
nucleon emerges as a soliton. They obtained significantly 
smaller values for $h_\pi \simeq 0.2 \cdot 10^{-7}$. In Ref. \cite{Meissner:1998pu}, the calculation was sharpened based
on a three-flavor Skyrme model calculation with the result $h_\pi \simeq 1 
\cdot 10^{-7}$ in agreement with a recent large $N_c$ analysis \cite{Phillips:2014kna}.
Recently, the first lattice QCD calculation \cite{Wasem:2011tp} has been performed for $h_\pi$ using a 
lattice size of $2.5\,\mathrm{fm}$ and a pion mass $M_\pi \simeq 
389\,\mathrm{MeV}$, finding the result
$h_\pi = \left(1.1\pm0.5\,(\mathrm{stat}) \pm 0.05 \,(\mathrm{sys})\right)\cdot 10^{-7},$
which is also rather small with respect to the DDH range. It should be stressed that this calculation does not include disconnected diagrams nor has it been extrapolated to the physical pion mass. 
The smaller estimates seem to be in better agreement with data.  Experiments 
on $\gamma$-ray emission from ${}^{18} F$ set the rather strong upper 
limit \cite{Haxton:1981sf}, $h_\pi < 1.3 \cdot 10^{-7}$,  although it must be stressed that calculations for nuclei with this many nucleons bring
in additional uncertainties. On the other hand, the Cesium anapole moment prefers a much larger value $h_\pi \simeq 10^{-6}$ although the involved uncertainties are also larger.  

Considering the uncertain status of the leading-order LEC $h_\pi$ it is clearly crucial to get a better handle on its size. In the first step, one therefore aims at  extracting $h_\pi$ from data on PV in proton-proton ($pp$) scattering. The observable we are interested in is the so-called longitudinal analyzing power (LAP)  which vanishes in the limit of $P$ conservation. It is defined as the difference in cross section between an unpolarized target and a beam of positive and negative helicity, normalized to the sum of these cross sections.  The LAP has been measured for several beam energies and results have been
reported for 
$13.6$~\cite{Eversheim:1991tg}, $45$~\cite{Kistryn:1987tq}, and $221$~\cite{Berdoz:2001nu} MeV 
(lab energy).   Traditionally, it was assumed that the $pp$ LAP does not depend on $h_\pi$ \cite{Carlson:2001ma} as the OPE potential does not contribute due to its isospin-violating nature. However, this is not true for the TPE potential. In Ref.~\cite{deVries:2013fxa} the goal was to include the TPE potential and extract $h_\pi$ from the $pp$ data. However, at the order of the TPE diagrams the $pp$ data also depends on one combination of PV short-range parameters that we defined as $C$. In particular, the goal was to perform a fully systematic analysis within chiral nuclear EFT.  
Technically the task is to solve the Lippmann-Schwinger (LS) equation in presence of a potential $V$ which is the sum of the strong, PV, and Coulomb potentials. For the strong potential  the N$^2$LO  potential \cite{Epelbaum:2004fk} was used, for various values of the cut-off appearing in the LS equation to get a handle on the theoretical uncertainty. We then fitted $h_\pi$ and $C$ to the $pp$ data. The data point at $221$ MeV corresponds to a transmission experiment which means that it is sensitive to small forward scattering angles where the Coulomb potential diverges. This difficulty has been carefully taking into account in the analysis and plays an important role in the extraction of the LECs.   Unfortunately, because only three data points exist with significant uncertainties, the fits allow a rather large range of parameters. 
 The allowed range for the  LECs, at the total $\chi^2 =2.71$ level, is approximately
$h_\pi = (1.1\pm 2 )\cdot 10^{-6},
C = (-6.5\pm 8)\cdot 10^{-6}$,
and the couplings are heavily correlated. The large uncertainty is dominated by experimental errors and the lack of data points.  Our findings have been confirmed in the analysis of Ref.~\cite{Viviani:2014zha}. Additional experiments are needed to reduce the uncertainties in the fits. One of the conclusions therefore is that a transmission experiment around $125$ MeV
would be very beneficial in reducing the uncertainties on the LECs. Such an experiment could be performed at
COSY at the  Forschungszentrum J\"ulich.  Unfortunately, due to the large allowed range of $h_\pi$ we cannot confirm nor rule out small values of $h_\pi$. However, the suggested smallness of $h_\pi$ by the ${}^{18}$F data and several theoretical estimates, indicate that possible higher-order corrections to the PV potential might be relevant. It was thus necessary  to derive these next-to-next-to-leading order corrections to the potential which might become crucial in understanding hadronic PV in case $h_\pi$ turns out to be very small.  The corrections have been systematically studied in Ref.~\cite{deVries:2014vqa} where it was found that five new LECs appear. These LECs describe new PV pion-nucleon and pion-pion-nucleon interactions. Only two combinations of these five LECs contribute to PV in $pp$ scattering. The combinations contribute via OPE and TPE diagrams, however, a detailed study indicates that the TPE diagrams are by far dominant and the OPE diagrams can be neglected. 
The TPE diagrams themselves can be divided into two parts. One part involves no new LECs and is proportional to $h_\pi\,c_4$, where $c_4\simeq 3.4$ GeV$^{-1}$ is a well-known PC LEC which takes on a large value due to underlying $\Delta$ and $\rho$ dynamics. The second part involves a combination of new LECs $h^{pp}_{\mathrm{TPE}}$. Due to the large size of $c_4$ we can expect the $ h_\pi\,c_4$ term to dominate the N${}^2$LO potential. Because this combination involves no new LECs compared to the NLO potential studied above, we have been able to refit $h_\pi$ to the $pp$ data by including the dominant N${}^2$LO correction. This leads to a slightly improved fit and a somewhat smaller value for $h_\pi$
 \begin{eqnarray}\label{fit4}
h_\pi = (0.8\pm 1.5 )\cdot 10^{-6}~,~~ C = (-5.5\pm 7)\cdot 10^{-6}~.
\end{eqnarray}
The N$^2$LO correction does not affect the values of the LECs by a large amount.  This indicates that the expansion of chiral EFT is converging well. 
Unfortunately, due to a lack of data it is not possible to extract a value for $h_\pi$ and $h^{pp}_{\mathrm{TPE}}$ at the same time. However, from an analysis of corrections proportional to $h^{pp}_{\mathrm{TPE}}$ one can  draw a few conclusions. Unless $h_\pi$ is very small, the N${}^2$LO TPE corrections are dominated by terms proportional to $h_\pi$. This would imply that the dominant part of the N${}^2$LO potential contains no new LECs. Finally, if it turns out that $|h_\pi| < 10^{-7}$, the N${}^2$LO corrections calculated in 
Ref.~\cite{deVries:2014vqa} might need to be included. 

Only additional PV data can tell us which of the above scenarios is realized in nature.
For example, a measurement of PV in the reaction $\vec n p\rightarrow d \gamma$ could shed light on the size of $h_\pi$. In contrast to $pp$ scattering, the longitudinal asymmetry $a_\gamma$ in this process does depend on the leading-order OPE potential and is therefore much more sensitive to $h_\pi$. For a long time, however, there was no non-zero measurement of $a_\gamma$. Recently a first preliminary result for $a_\gamma$ was reported \cite{Crawford}
\begin{equation}\label{agammaEXP}
a_\gamma = (-7.14\pm4.4)\cdot 10^{-8}\,\,\,.
\end{equation}
This result is based on a subset of the full data taken and 
an improved result with an uncertainty at the $10^{-8}$ level is expected in the near future.
The goal of Ref.~\cite{deVries:2015pza} was to to combine the above-described analysis of $pp$ scattering with this recent data on $\vec n p \to d \gamma$ in order to extract a value of $h_\pi$. In contrast to $pp$ scattering, the analysis of $\vec n p \to d \gamma$ requires the inclusion of PC and PV electromagnetic currents. These can be derived from chiral EFT 
in the same way as the potentials have been derived. The PC currents that we included in the analysis arise from the nucleon magnetic moments, a recoil correction to the nucleon charges, and currents associated with a single pion exchange. With these currents we calculate a $P$-even cross section within $3\%$ of the experimental value.
The PV currents arise from a single pion exchange where one of the vertices is $h_\pi$. Summing up all LO contributions, we find that each of the individual contributions only has a minor dependence on the cut-off value, but the sum suffers from a larger dependence due to mutual cancellations. In total we find
\begin{equation}
a_\gamma = -(0.11\pm0.05)\,h_\pi\,\,\,.
\end{equation}
We can now combine the $a_\gamma$ analysis with that of  $pp$ scattering. 
Including the theoretical uncertainty from cut-off variations we obtain the following ranges for the LECs
\begin{equation}\label{eq:hpi}
h_\pi = (1.1 \pm 1.0)\cdot 10^{-6}\,\,\,,\quad C = (-6.5\pm 4.5)\cdot 10^{-6}\,\,\,.
\end{equation}
The fits indicate that small values of $h_\pi \sim 10^{-7}$ are barely consistent with the data, with values of $h_\pi\sim(5-10)\cdot 10^{-7}$ being preferred.  Such larger values disagree with the upper limit from ${}^{18}$F gamma-ray emission, $h_\pi \leq 1.3\cdot 10^{-7}$, and lattice and model calculations of $h_\pi \simeq 10^{-7}$. The upcoming increase in sensitivity of the $a_\gamma$ measurement will significantly improve the fit and tell whether small values of $h_\pi$ are consistent with few-body experiments.
We have estimated the uncertainties of the fits due to 
experimental uncertainties, variation of cut-off parameters in the LS equation, and higher-order corrections 
and find the first of these to be dominant.

All this work  has thus significantly advanced the understanding of 
hadronic and nuclear parity violation. The application of chiral effective field 
theory to this longstanding program has proven to be crucial in combining the 
various processes in a single unifying framework. The Pisa group has picked up our
approach and applied it to various PV obervables in systems with three and four nucleons,
such as $\vec n$-$d$ scattering or the charge exchange reaction $^3$He$(\vec n ,p)^3$H
\cite{Viviani:2014zha}. So we can summarize the status of $h_\pi$ as follows: 
If the upcoming data on $\vec n p\rightarrow d\gamma$ confirms the preliminary 
number, Eq.~(\ref{agammaEXP}),  this strongly indicates that $h_\pi$ 
has a value significantly larger than the upper bound from ${}^{18}$F data. 
There are then two options. Either $h_\pi$ is actually that small and  higher-order 
corrections calculated already can explain the large value of $a_\gamma$. 
This is, however, not a satisfying explanation, as this requires higher-order LECs 
that are significantly larger than expected from the power counting and 
resonance saturation. The other option is that something is missing in the 
theoretical analysis of the ${}^{18}$F data. Nuclear lattice simulations will eventually be able 
to do a clear-cut calculation of the parity-mixing in ${}^{18}$F -- stay tuned.

Next to the study of parity violation,  the closely related subject of hadronic and nuclear CP violation has been
investigated in the last years.  The SM contains two sources of CP violation (CPV), one in the phase 
of the quark-mass matrix and one in the strong interactions, the QCD $\theta$ term. The former 
manifests itself in CPV flavor-changing interactions and leads only to very small electric dipole 
moments (EDMs). On the other hand, the QCD $\theta$ term is flavor conserving and gives rise 
to an, in principle, large neutron EDM. As noted before, the non-observation of the latter then forces $\theta \leq 10^{-10}$. 
This extreme smallness is known as the strong CP problem. In addition, EDMs can obtain contributions 
from physics beyond the SM. In fact, large EDMs are generated in various popular 
extensions of the SM such as supersymmetric and left-right symmetric models. The extreme 
accuracy of low-energy EDM measurements probe high-energy scales comparable to the Large Hadron Collider. 
The above considerations have led to a large experimental endeavour to measure EDMs
of leptons, hadrons, nuclei, atoms, and molecules  \cite{Engel:2013lsa}.  The main motivation 
for our work in this field are the plans to measure the neutron EDM with higher accuracy and to 
measure for the first time the EDMs of  light nuclei in storage rings. It has been
proposed that storage rings can be used to measure the EDMs of the proton and deuteron
with a precision of $10^{-29}\,e\,$cm, three orders of magnitude better than the current neutron
EDM limit. EDMs of other light ions, such as the helion (${}^3$He nucleus), are candidates as well.
Any finite signal in one of the upcoming experiments would be due to physics not accounted
for by the phase in the quark-mass matrix. Such a signal would either be caused by physics beyond the SM
(BSM) or by an extremely small, but nonzero, $\theta$ term. An interesting and important problem is therefore to 
investigate whether it is possible to trace a nonzero $\theta$ with EDM experiments. That is, can we 
confidently disentangle the $\theta$ term from possible BSM sources? As will be shown, some progress
has been made to answer this question.

 Let us first stay within the SM. Once the QCD Lagrangian is 
supplemented by a nonzero QCD $\theta$ term, CPV interactions between the low-energy degrees of 
freedom appear. Since the $\theta$ term breaks chiral symmetry like the quark masses, CHPT can be easily 
extended to include such interactions. In particular, this extension gives rise to CPV 
couplings between pseudo-Goldstone bosons (pion, kaon, eta) and baryons (in particular nucleons) 
whose strengths can be related to known baryon mass splittings and sigma terms 
\cite{Crewther:1979pi,Borasoy:2000pq}. It then becomes possible to calculate the EDM of the 
nucleon (and heavier baryons) with CHPT \cite{Ottnad:2009jw,Guo:2012vf}. The divergences appearing 
in these loops are absorbed by counter terms whose sizes cannot be obtained from CHPT directly. 
Lattice-QCD simulations and nonphysical quark masses and including a nonzero $\theta$ term have 
been performed to calculate these unknown counter terms. By using the CHPT expressions of the 
nucleon EDMs \cite{Ottnad:2009jw,Guo:2012vf}, the results of the simulations can be extrapolated 
to the physical point and infinite volume \cite{Akan:2014yha}. In this way, the proton and 
nucleon EDM have been calculated in terms of $\theta$ directly \cite{Guo:2012vf,Guo:2015tla}
\begin{equation}
d_n =  - (3.8\pm1.0)\cdot 10^{-16}\,e\,\mathrm{cm}\, ,~~ 
d_p =   (2.1\pm1.2)\cdot 10^{-16}\,e\,\mathrm{cm}\, .
\end{equation}
These calculations provide an important contribution to the study of hadronic and nuclear EDMs. 
Once nonzero nucleon EDMs are measured the above results can be used to test whether the $\theta$ 
term is responsible or some other source of CPV.

Considering the success of the SM of particle physics, it is likely that any additional 
physics appears at a scale considerably higher than the electroweak scale $\sim 100$ GeV. 
This scale separation makes it possible to treat the SM as the dimension-four and lower part of 
a more general EFT containing higher-dimensional operators. For the study of EDMs it can be 
shown that the first operators appear at dimension six and are suppressed by two powers of 
the scale where the additional CPV appears. BSM CPV can be studied in a model-independent way 
by adding all possible CP-odd dimension-six operators at the high-energy scale. The great advantage 
is that it is not necessary to choose a specific SM extension. Nevertheless, as discussed below, 
the approach can be matched to specific high-energy models to study their low-energy consequences. 
EDM experiments take place at very low energies such that the dimension-six operators must be 
evolved to lower energies 
taking into account QCD and electroweak renormalization-group evolution \cite{Dekens:2013zca}. 
Once the dust settles, only a relatively small set of operators remain at a scale $\sim 1$~GeV 
consisting of quark EDMs and chromo-EDMs, the gluon chromo-EDM, and several four-quark operators. 
At lower energies, QCD becomes nonperturbative and to proceed further we have extended CHPT to 
include the dimension-six operators. The resulting CHPT Lagrangians have been built in great 
detail in Refs.~\cite{deVries:2012ab,Bsaisou:2014oka}.  All dimension-six operators (and the 
$\theta$ term) break CP symmetry, however they all break chiral symmetry in different ways 
leading to different CHPT interactions. The different interactions, in turn, lead to different 
hierarchies of EDMs. Thus, given enough measurements it becomes possible to unravel the underlying 
source of CPV. This hierarchy of EDMs can be best studied for the EDMs of light nuclei. 
In principle the nucleon EDM induced by BSM sources can be calculated within CHPT in the same 
way as for the $\theta$ term. However, in contrast to the $\theta$ term, the associated counter terms 
that appear have not been calculated with lattice QCD. Such calculations are significantly more 
difficult than for $\theta$. Furthermore, even if they had been calculated, measuring only 
nucleon EDMs will not be enough to unravel the sources. 

The plans to measure the EDMs of light nuclei in storage rings with high accuracy make it 
attractive to focus on these observables. Similar, to the research program on parity violation 
described above, this can be done by combining CP-even chiral EFT potentials with CPV potentials 
that are calculated for each possible source of CPV. The  power counting shows that EDMs of light 
nuclei are dominated by a small set of hadronic interactions. These interactions are the EDMs of the 
constituent nucleons ($d_n$ and $d_p$), two CPV pion-nucleon couplings ($g_0$ and $g_1$), one CP-odd 
three-pion vertex ($\Delta$), and two CPV nucleon-nucleon couplings ($C_1$ and $C_2$) \cite{Bsaisou:2014zwa}.
For instance, the deuteron EDM $d_D$ at N$^2$LO is found to be~\cite{Bsaisou:2012rg} 
\begin{eqnarray}
d_D &=& 0.9 \,d_n+ 0.92\, d_p \nonumber\\
&-&\left[(0.18\pm 0.02) g_1+(0.75\pm0.15)\,\Delta\right]\,e\,\mathrm{fm}\, .
\end{eqnarray}
Note that the leading CPV  4N couplings do not contribute here.
For sources such as quark chromo-EDMs or certain four-quark operators, the contributions from 
pion exchange can be significantly larger than the single nucleon EDMs. On the other hand, for the 
$\theta$ term, due to its isospin-conserving nature, the contributions from $\bar g_1$ and $\Delta$ 
are only a fraction of the neutron EDM
\begin{equation}
d_D - 0.9 \,d_n- 0.92\, d_p = -(0.9\pm0.3)\cdot 10^{-16}\,\theta\,e\,\mathrm{cm} \ll d_n\,\,\,.
\end{equation}
These calculations show that measurements of both the neutron and the deuteron EDM can provide 
strong hints for BSM physics. A large hierarchy between $d_D$ and $d_n$ would point towards sources of CPV not 
in the SM. 
The above calculations have been extended to the EDMs of the triton and helion. In particular 
the latter is interesting as it can be probed in a storage-ring experiment as well. 
The calculations show that the ${}^3$He EDM is, in contrast to $d_D$, also sensitive to 
$g_0$ and therefore complementary. In Ref.~\cite{Dekens:2014jka} a comprehensive study of the 
EDM signature of several popular BSM models was performed. It was shown that the measurements 
of the EDMs of a few light nuclei could be enough to unravel several high-energy BSM models. 
This study shows that low- and high-energy searches for BSM physics are complementary and 
that EDM measurements are able to probe the highest energy scales. 

\smallskip

\begin{figure}[t]
\begin{center}
\includegraphics[width=.485\textwidth]{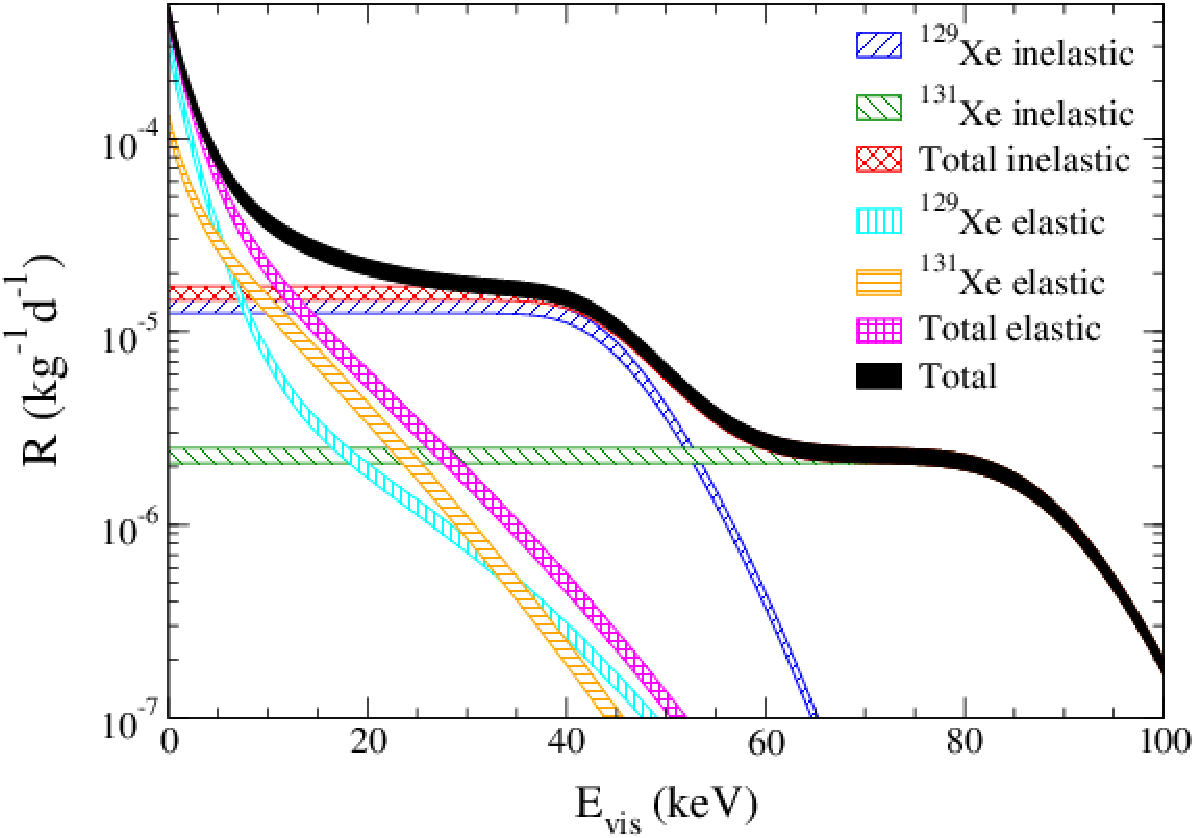}
\end{center}
\caption{Integrated energy spectra  for
elastic and inelastic, spin-dependent scattering 
of a WIMP off Xe assuming couplings to the neutrons only,
assuming a cross section off the nucleon,
$\sigma_N = 10^{-40}\,{\rm cm}^2$.
The differential spectra  are  integrated  from  the  threshold  value
$E_{\rm vis}$ to infinity. The  bands result from 
uncertainties in one- and two-body currents.
Figure courtesy of Achim Schwenk. }
\label{fig:DM}
\end{figure}

As the last topic of this section I discuss recent work by the Darmstadt group, that has made considerably
progress in the calculations for dark matter particles/WIMPs\footnote{WIMP stands for Weakly Interacting Massive Particle.
For an introduction to the field of dark matter searches  and possible candidate particles, see e.g.~Ref.~\cite{pdgDM}.} 
scattering off nuclei, utilizing state-of-the-art nuclear structure
calculations combined with currents derived from chiral nuclear EFT. First, they worked out the structure
factors for elastic spin-dependent WIMP scattering off nuclei relevant to dark matter detection experiments,
namely $^{129}$Xe,  $^{131}$Xe,  $^{127}$I,  $^{73}$Ge,  $^{19}$F,  $^{23}$Na,  $^{27}$Al  and $^{29}$Si.
For the first time, the spin-dependent WIMP-nucleus currents were based on chiral EFT, and uncertainty bands
due to nuclear uncertainties where supplied \cite{Klos:2013rwa}.  An important further step was taken in 
Ref.~\cite{Baudis:2013bba}, where inelastic scattering was explored. It is assumed that the dark matter particle
excites the nucleus to a low-lying state with an excitation energy of 10-100~keV followed by a prompt de-excitation.
It is found that for momentum transfers of the order of the pion mass, which can typically be reached in such
processes, the inelastic channel is comparable or can even dominate the elastic one. This can have a very
distinct effect on the integrated spectra as shown in Fig.~\ref{fig:DM}. Instead of the expected exponential 
fall-off from the elastic reaction, one observes a double-plateau structure, depending of course on the mass
of the dark matter particle and other assumptions specified in Ref.~\cite{Baudis:2013bba}.  The precise
location of these plateaus will thus allow one to constrain the mass of the dark matter particle scattering off the
nucleus. Matters are different for spin-independent WIMP scattering off Xe, where the structure factors for
inelastic scattering are suppressed by about four orders of magnitude compared to the coherent elastic
response~\cite{Vietze:2014vsa}. Finally, in Ref.~\cite{Hoferichter:2015ipa} a power counting scheme for 
scalar, pseudoscalar, vector and axial-vector WIMP-nucleon interactions was presented and all one- and
two-body currents to third order in the chiral expansion were derived. It is also shown that chiral symmetry 
predicts a hierarchy between the various operators. Further, the relevance of two-body currents is stressed,
which is nothing but a reflection of the importance of meson-exchange currents in the nuclear response to
external probes.  The intriguing field is certainly only at its beginning and more work on improving the nuclear
structure aspects is called for.


\section{Perspectives}
\label{sec:out}

Nuclear physics has entered a new area and is now firmly rooted in the underlying 
gauge theory of the strong interactions, QCD. This is a remarkable achievement
since after the Nobel prize to Bohr, Mottelson and Rainwater, many had announced
the end of nuclear physics. To the contrary, we are just at the beginning of an
exciting period in nuclear physics research with many intriguing results to be expected.
Here, I list a few activities that I believe will become more important in the 
years to come (this list is very incomplete, issue like strangeness nuclear physics,
nuclear reaction studies and much more are not even mentioned):
\begin{itemize}
\item
Three- and four-nucleon forces need to be scrutinized in light- and medium-heavy
nuclei along the lines laid out in Ref.~\cite{Wienholtz:2013nya},
using, however, improved many-body techniques. Nuclear lattice simulations
appear to be the method of choice here.
\item
Nuclear lattice EFT has to be pushed to higher $A$ and more neutron- or proton-rich
nuclei and more work on reactions along the lines of 
Refs.~\cite{Rupak:2013aue,Pine:2013zja,Elhatisari:2015iga} is called for.
\item
The response of nuclei to external probes requires more detailed investigations
of the underlying currents that have to be constructed in harmony with the
nuclear forces. This is a non-trivial exercise, as renormalizability poses
severe constraints that can most easily be accounted for using tailor-made
unitary transformations, see e.g. Refs.~\cite{Kolling:2009iq,Kolling:2011mt}.
For a status review, the reader is referred to Ref.~\cite{Bacca:2014tla} and
recent work on the axial currents is found in~\cite{Baroni:2015uza}.
\item
Lattice QCD attempts to derive nuclear properties from the underlying quark and
gluon degrees of freedom. This is a very ambitious but potentially  very rewarding
program. In my opinion, this will
require much more work and time. Clearly, one would like to see calculations at or close
to the physical quark masses, as it has become available in the meson and baryon sector.
\end{itemize}


\section*{Acknowledgements}

I thank all my collaborators for sharing their insights into the topics discussed
here, and in particular  V\'eronique Bernard, Evgeny Epelbaum  and Dean Lee for a careful reading. I am also grateful to 
Jerzy Dudek to giving me this opportunity to contribute
to this special edition. This work is supported in part
by DFG and NSFC through funds provided to the Sino-German CRC 110
``Symmetries and the Emergence of Structure in QCD'' (NSFC Grant No. 11261130311),
by the Chinese Academy of Sciences (CAS) President's International Fellowship 
Initiative (PIFI) (Grant No. 2015VMA076), and by the HGF through funds provided 
to  Virtual Institute NAVI (Contract No. VH-VI-417).


\end{document}